\documentclass[aps,prl,groupedaddress,showpacs]{revtex4-1}
\usepackage[dvipdfmx]{graphicx}
\usepackage{color}

\usepackage{amsmath}
\usepackage{bm}

\usepackage{caption}

\renewcommand{\figurename}{\bf Figure}

\begin{document}
\title{Formation of Dominant Mode by Evolution in Biological Systems}

\topmargin 0mm

\renewcommand{\thefootnote}{\fnsymbol{footnote}}

\author{Chikara Furusawa}
\email[]{chikara.furusawa@riken.jp}
\affiliation{
Quantitative Biology Center (QBiC), RIKEN, 6-2-3 Furuedai, Suita, Osaka 565-0874;\\ Universal Biology Institute, University of Tokyo, 7-3-1 Hongo, Tokyo 113-0033, Japan}

\author{Kunihiko Kaneko}
\email[]{kaneko@complex.c.u-tokyo.ac.jp}
\affiliation{Research Center for Complex Systems Biology, Universal Biology Institute, University of Tokyo, 3-8-1 Komaba, Tokyo 153-8902, Japan}

\begin{abstract}
\textcolor{black}{A reduction in high-dimensional phenotypic states to a few degrees of freedom is essential to understand biological systems. One possible origin of such a reduction (as recently discussed) is the steady growth of cells that constrains each component's replication rate. Here, in contrast, our aim is to investigate consequences of evolutionary robustness, which is shown to cause a stronger dimensional reduction in possible phenotypic changes in response to a variety of environmental conditions. First, we examined global protein expression changes in {\it Escherichia coli} after various environmental perturbations. Remarkably, they were proportional across components, across different types of environmental conditions, while the proportion coefficient corresponded to the change in growth rate.} Because such global proportionality is not generic to all systems under a condition of steady growth, a new conceptual framework is then needed. We hypothesized that such proportionality is a result of evolution. To test this hypothesis, we analyzed a cell model---with a huge number of components, that reproduces itself via a catalytic reaction network---and confirmed that common proportionality in the concentrations of all components is shaped through evolutionary processes to maximize cell growth (and therefore fitness) under a given environmental condition. Furthermore, we found that the changes in concentration across all components in response to environmental and evolutionary changes are constrained to the changes along a one-dimensional major axis within a huge-dimensional state space. On the basis of these observations, we propose a theory in which high-dimensional phenotypic changes after evolution are constrained to the points near a one-dimensional major axis that correlates with the growth rate, to achieve both evolutionary robustness and plasticity. By formulating this proposition in terms of dynamical systems, broad experimental and numerical results on phenotypic changes caused by evolution and adaptation are coherently explained.
\end{abstract}

\maketitle

\noindent
{\bf Summary:}\\
Cells generally consist of thousands of components whose abundance levels change through adaptation and evolution. Accordingly, each steady cell state can be represented as a point in a high-dimensional space of component concentrations. In the context of equilibrium statistical thermodynamics, even though the state space is high-dimensional, macroscopic description only by a few degrees of freedom is possible for equilibrium systems; however, such characterization by a few degrees of freedom has not yet been achieved for cell systems. Given that they are not in equilibrium, we need some other constraint to be imposed. Here, by restricting our focus to a cellular state with steady growth that is achieved after evolution, we examine how the expression levels of its several components change under different environmental conditions. On the basis of analysis of protein expression levels in recent bacterial experiments as well as the results of simulations using a toy cell model consisting of thousands of components that are reproduced by catalytic reactions, we found that adaptation and evolutionary paths in high-dimensional state space are constrained to changes along a one-dimensional curve, representing a major axis for all the observed changes. Moreover, this one-dimensional structure emerges only after evolution and is not applicable to any system showing steady growth. This curve is determined by the growth rate of a cell, and thus it is possible to describe an evolved system by means of a growth rate function. All the observed results are consistent with the hypothesis that changes in high-dimensional states are nearly confined to the major axis in response to environmental, evolutionary, and stochastic perturbations. This description opens up the possibility to characterize a cell state as a macroscopic growth rate, as is the case for the thermodynamic potential. This approach can provide estimates of which phenotypic changes are theoretically more evolvable, as predicted simply from their observed environmental responses.

\section{Introduction}

Cells contain a huge variety of components whose concentrations change through adaptation and evolution. To realize a theoretical description of such a system with many degrees of freedom, it is important to be able to characterize the system based on a few macroscopic variables, as demonstrated by the success in traditional statistical thermodynamics. In thermodynamics, by restricting the focus to an equilibrium state and its vicinity, a state can be described by a few macroscopic variables even though the number of microscopic degrees of freedom is huge. Accordingly, we sought to evaluate the possibility of adopting a similar approach to a biological system that contains a large number of components.

Of course, the strategy of statistical thermodynamics cannot be directly applied to a biological system given that cells are not in a thermodynamic equilibrium. Cells grow (and divide) over time in a manner that is driven by complex intracellular dynamics. Despite this difference, however, it may still be possible to establish a description of cells with only a few degrees of freedom by restricting our focus to only cells in a steady growth state. Indeed, for such cases, all the intracellular components have to be roughly doubled before cell division occurs for reproduction. This restriction imposes a general constraint on the possible changes in expression levels across the many components; this situation opens the possibility to characterize a cellular state with a few variables. Indeed, several research groups have attempted to search for universal laws in the steady growth state, as pioneered by the work of Monod \cite{Monod} followed by subsequent studies \cite{Pirt, Scott-Hwa,Klumpp,Dill}.

Currently, research on the existence of such constraints is generally carried out by examining changes in expression occurring across thousands of genes in transcriptome or proteome data. In fact, a general trend across changes in the overexpression of many components has been observed \cite{Marx,Pancaldi,Alon,Braun,Matsumoto} in the course of adaptation and evolution \cite{Lopez-Maury,Barkai,Bahler,Lehner,Horinouchi1,Horinouchi2}. In particular, using transcriptome analysis of bacteria, we recently found that changes in (logarithmic) expression levels are proportional across many components under given stress conditions of different intensities, and the proportion coefficient corresponds to the change in the growth rate \cite{Mu,CFKK-Interface}. 
\textcolor{black}{To account for the experimental findings, a theory was proposed based on steady growth of cells and linear approximation under the assumption of small changes in expression levels. 
The theory can explain the proportional change in expression levels across many components, in response to different intensities of the {\it same} environmental stress, as long as the change is {\it small}.
Experimental data, however, suggest that such proportionality in the expression levels across many genes also holds across different types of environmental conditions. 
Just as the previous theory concerned with the change only along a given single environmental change (i.e., changes characterized only by one parameter), it cannot explain the observed global proportionality across different directions in environmental changes. 
Furthermore, the experimental data suggest that the proportionality in expression levels is valid even in the case of substantial stress intensity, leading to a decrease down to 20\% in the growth rates.}  

\textcolor{black}{To sum up, the data, albeit preliminary, point to some novel theory that is not explained simply by the steady growth. 
Therefore, we first need experimental confirmation for the existence of such global proportionality across a variety of environmental conditions. If this confirmation is obtained, then we will seek a general theory beyond the simple steady growth.}

In the present study, we first analyzed the changes in protein expression levels under a variety of environmental conditions using recent experimental data. We confirmed the pattern of proportional changes across all protein expression levels, with the proportion coefficient approximately consistent with the cell growth rate, representing a basic macroscopic quantity for steady-growth cells. \textcolor{black}{The results are not explained by the previous simple theory because they deal with the proportionality across a broad range of different environmental conditions. It is not a generic property corresponding to simple steady growth and requires a novel theory to take into account some other factor.} 

At this stage, it is important to recall that biological systems are an outcome of evolution. Hence, it is natural to expect that the missing link may be provided as a result of evolution under a given environmental condition. Accordingly, we here examine validity of the hypothesis that evolution under a given condition will constrain intracellular dynamics so that the concentration changes across many components are mainly governed by changes in the growth rate.
Extensive evolutionary experiments are needed to test this hypothesis, and this work may be too difficult to do in a laboratory. Nonetheless, a numerical evolution approach offers a possible way to address this question. 

In fact, several models have been proposed that capture the evolution of a cellular state with many phenotypic variables. Although such models may not be able to describe the entire breadth of the complexity of cells, they are sufficiently complex to enable interpretations on the evolution of genotype--phenotype mappings \cite{Geno-Pheno,Geno-Pheno2,Wagner,KK-PLoS,Soyer}. Here, we study a cell model consisting of a large number of protein species that catalyze each other's reactions transforming environmental resources for cellular growth. We demonstrate that after evolution, the changes across all catalysts are constrained to points near a one-dimensional curve that is governed by the growth rate. 

According to these findings of numerical simulations, with support from laboratory experiments, we formulate the following hypothesis of an evolved system: phenotypic changes of many components after environmental or genetic alterations mainly take place along a single major axis that parameterizes the growth rate. Indeed, this theory can explain the results of phenotypic changes occurring by adaptation and evolution quite well. We then discuss the general consequences and biological relevance of this hypothesis.

\section{Global proportionality observed in protein expression levels}

\color{black}
By imposing linearization approximation with steady growth, 
we previously considered how each concentration of each intracellular component $x_j$, say mRNA expression level measured by transcriptome analysis, changes when the intensity of a given environmental stressor increases. Specifically, we are concerned with the logarithmic change in concentration $x_j$, i.e., $\delta X_j(E)\equiv X_j(E)-X_j(O)$, where $O$ is the original nonstress condition. By assuming the steady growth of cells and linearizing all the dynamics around the original state, expression change $\delta X_j(E)$ is obtained in response to the change in single scalar parameter $E$ representing the value of a given single environmental condition (e.g., temperature change). From the steady-growth solution, it then follows that change $\delta X_j$ across all components $j$ in a cell under a given environmental (stress) condition, 
\begin{math}
\frac{ \delta X_j( E)}{ \delta X_j( E')}= \frac{\delta\mu(E)}{\delta\mu(E')},
\end{math}
where $\mu$ is the growth rate of a cell in the steady-growth state, and $E$ and $E'$ are two intensities of a given single environmental stressor.
Because the right-hand side is independent of each component $j$, the above equation implies that the changes in all components are proportional, with the slope given by the growth rate change. 
By means of the transcriptome analysis of {\sl Escherichia coli}, the relation is found to hold rather well for the change in mRNA abundance under given stress types \cite{Mu}.

So far, the environmental condition was a given scalar variable. Nevertheless, there is a variety of environmental conditions. Then, is this relation valid not simply toward the intensity of a given stressor but across a variety of environmental conditions? In other words, for different types (i.e., for different vectors ${\bf E_1}$ and ${\bf E_2}$, does the relation 

\begin{equation}
\frac{ \delta X_j({\bf E_1)}}{ \delta X_j({\bf E_2)}}= \frac{\delta\mu({\bf E_1)}}{\delta\mu({\bf E_2)}}
\end{equation}  
hold?
This is not an incremental question, because such a relation, if it exists, cannot be explained simply by the steady growth but reflects some other basic property inherent in (evolved) biological systems, which necessitates a novel conceptual framework leading to a strong constraint in possible phenotypic changes.

To also allow for a direct comparison with protein expression dynamics, we examined the proteome data of growing {\it E. coli} under various environmental conditions reported in ref. \cite{Heinemann}. The environmental conditions considered include different culture media with various carbon sources, application of stressors such as high temperature and osmotic pressure, and glucose-limited chemostat cultures with different dilution rates (which are equivalent to specific growth rates in the steady state). Measurement of intracellular protein concentrations was carried out by mass spectrometry, which provided data on the absolute concentrations of more than a thousand protein species. 

\color{black}
From the proteomic analysis under various environmental conditions, we calculated the change in protein concentrations between the standard (reference) state and that under different environmental conditions. In the following analysis, a minimal medium with glucose as a sole carbon source was selected as the standard state, and essentially the same results were obtained when other data served as the standard state to compute changes $\delta X_j({\bf E})$ and $\delta \mu$. The log-transformed expression change of the $j$-th protein is calculated as $\delta X_j({\bf E})=\log (x_j({\bf E})/x_j({\bf S}))$, where $x_j({\bf E})$ and $x_j({\bf S})$ represent the protein concentration in given environment ${\bf E}$ and the standard condition, respectively. Here, we employ the logarithmic scale because protein expression levels typically change on this scale \cite{Elowitz,Log-normal,Bar-Even,Sato}, and this choice facilitates a comparison with the theory described above and discussed below. 

To determine whether the common proportionality of protein expression changes in Eq. (1) is also valid across different environmental conditions, we analyzed the relation of protein concentration changes $\delta X_j$ observed in two different environmental conditions. In Fig. 1(a) and (b), we show some examples of the relation between protein expression changes $\delta X_j({\bf E_1})$ and $\delta X_j({\bf E_2})$, where ${\bf E_1}$ and ${\bf E_2}$ are two different culture conditions.
\textcolor{black}{Using the data, we plotted $(\delta X_j({\bf E_1}),\delta X_j({\bf E_2}) )$ for all pairs of 20 different environmental conditions $({\bf E_1},{\bf E_2})$.} 
As shown in the figure, the expression changes under different environmental conditions strongly correlate over a large number of proteins. \textcolor{black}{By means of the plots, we computed the slope of the common trend in protein expression changes, which is in agreement with that calculated from the ratio of growth rate changes as in Eq. (1) (blue lines in Fig. 1(a) and (b)). In Fig. 1(c), we compared the slope of the common trend in protein expression obtained by linear regression, i.e., a comparison of 
common ratio $\delta X_j({\bf E_1}) / \delta X_j({\bf E_2})$ with the ratio of growth rate change $\delta \mu({\bf E_1}) / \delta \mu({\bf E_2})$, for all possible combinations by choosing a standard condition (see the caption of Fig. 1 for details) \cite{footnote0}.}

\color{black} Here, we employ proteomics data, whereas another possible choice will be the use of transcriptome data. Either is fine for our purpose, and in fact some analysis involving transcriptome data also indicates such global proportionality in expression levels \cite{Marx,Pancaldi,Alon,Braun,Matsumoto} although the relation with the growth rate is not explored. One advantage of the proteomic analysis over the transcriptomic analysis is that the former reflects post-transcriptional regulation, which is believed to be involved in the growth rate regulation \cite{Castrillo}. In fact, the proteome data we are dealing with \cite{Heinemann} have widely been used to analyze growth rate--dependent regulation of protein expression levels and resource allocation \cite{O'Brien,Ebrahim,Noor,Barenholz,Yang}. Thus, application of the proteome data will be more appropriate to analyze the relation between high-dimensional phenotypic dynamics and the growth rate in Eq. (1).

\color{black}The present global proportionality toward {\em various} environmental conditions is much stronger than in the earlier results and is beyond the scope of the simple theory based only on steady growth \cite{Mu} (and linear approximation), which can account only for the change along a {\em given single} environmental condition.
 There, $E$ and $E'$ represented the same environmental stress and differed only in the intensity, for example, a change in the culture temperature. They changed along the same vector. In contrast, the present proportionality deals with different types of stress conditions (e.g., temperature versus a nutrient), i.e., with vectors of different directions. Therefore, the observed proportionality cannot be derived from the original theoretical formulation and is not a generic property of the simple steady growth. Nevertheless, the data suggest that the proportionality in Eq. (1) is still valid in this case to a certain degree. 

\color{black}
Aside from this unexpected proportionality across different dimensions, the obtained data are suggestive of a broad range of linearity. The reaction or expression dynamics are generally highly nonlinear, with catalytic reactions or expression dynamics showing high Hill coefficients. Nonetheless, the experimental data still satisfy the linear relation even under a stress condition that reduces the growth rate below $20\%$ or so as compared to the standard. \textcolor{black}{Such expansion of the linear regime is not a general property of a system with steady growth either.}
 
To sum up, other factors aside from the steady growth are involved in order for the relation in Eq. (1) to hold across a variety of environmental conditions up to their large change.
Now, recall that biological systems not only are constrained by steady growth but also are a product of evolution. With evolution, cells can efficiently and robustly reproduce themselves under a given external condition. Thus, it is important to examine validity of the hypothesis that the above-mentioned global proportionality is a result of, and further strengthened by, evolution. 

\section{Catalytic-Reaction Network Model for Numerical Evolution}

To validate the above hypothesis, we evaluated how reaction (or protein expression) dynamics are shaped by evolution. To this end, we utilize a simple cell model consisting of a large number of components and numerically evolve it under a given fitness condition to determine how the phenotypes of many components evolve. Here, phenotypes are given by the concentrations of chemicals, which change due to catalytic chemical reactions of enzymes that are in turn synthesized by some other catalytic reactions.
Here, genes provide a rule for the possible reactions such as the parameters and structure of a catalytic-reaction network.

The phenotype of each organism as well as the growth rate (fitness) of a cell is determined by such reaction dynamics, while the evolutionary process consists of selection according to the associated fitness and genetic change in the reaction network (i.e., rewiring of the pathway). In the analysis of a population of such cells that slightly differ in their genotypes, and accordingly in their relative fitness, the offspring are generated in accordance with the fitness value. Through these dynamics, the evolutionary changes in the genotypes and corresponding phenotypes are traced, and the evolution of phenotypic changes is studied by means of this combination of dynamical systems and a genetic algorithm, which enables testing of the hypothesis at hand. 

The specific cell model applied for this purpose consists of $k$ components. Thus, the cellular state is represented by the numbers of $k$ components $(N_1,N_2,\cdots,N_k)$ and their concentrations $x_i = N_i /V$ with the volume of the cell $V$ \cite{Zipf,SOC-Zipf}. There are $m (<k)$ resource chemicals $S_1,S_2,\cdots,S_m$ whose concentrations in an environment and a cell are given by $s_1,\cdots,s_m$ and $x_1,\cdots,x_m$, respectively. 
The resource chemicals are transported into the cell with the aid of other chemical components named ``transporters.'' The other chemical species work as catalysts, which are synthesized from other components, including resource chemicals, in catalytic reactions driven by some other catalysts, while the resource and transporter chemicals are noncatalytic. 
With the transport of external resources and their conversion to other components, the cell increases its volume and divides in two when it reaches a size larger than a given threshold. 
The growth rate depends on the catalytic network, i.e., genotype. 

We adopted this model of a catalytic reaction network for simplicity because this model shows power law statistics, log-normal fluctuation, and adaptation with fold change detection; these characteristics are consistent with properties of cells \cite{Zipf,SOC-Zipf}. Detailed description is given below.

\color{black}
{\bf Catalytic reaction dynamics}:
For intracellular reaction dynamics, we consider a catalytic network among these $k$ chemical species, where each reaction leading from some chemical $i$ to some other chemical $j$ was assumed to be catalyzed by a third chemical $\ell$, i.e., $(i + \ell \rightarrow j + \ell)$.
In the zeroth generation, the catalytic network was generated randomly, where the path probability that a reaction from chemical $i$ to some other chemical $j$ exists is given by $\rho$. For simplicity's sake, all reaction coefficients were set to be equal. 

The nutrient chemicals (that are not catalysts) $S_1,..,S_m$ are transported into the cell with the help of other chemicals called ``transporters.'' 
Here, we assumed that the uptake flux of nutrient $i$ from the environment is proportional to $D s_i x_{t_i}$, where chemical $t_i$ acts as the transporter for nutrient $i$, and $D$ is a transport constant; in addition, for each nutrient, there is one corresponding transporter, represented by $t_i = m+i$.
Both the nutrients and transporters have no catalytic activity, whereas the other $k-2m$ chemical species are catalysts. Via catalytic reactions, these nutrients are transformed into other chemicals, including the transporters.

{\bf Cell growth}:
With the uptake of nutrient chemicals from the environment, the total number of chemicals
$N= \sum_i N_i$ in a cell can increase; consequently, we assumed that cell volume $V$ is proportional to total number $N$. 
For simplicity, cell division is assumed to happen when the cell volume exceeds a given threshold.
Through cell division, the chemicals in the parent cell are evenly distributed between the two daughter cells.

{\bf Reaction Procedure}:
In our numerical simulations, we randomly picked up a pair of molecules in a cell and transformed them if they are a substrate and the corresponding catalyst in the reaction network. In the same way, a nutrient molecule is transported across the plasma membrane if a randomly chosen intracellular molecule is a corresponding transporter for a randomly chosen extracellular nutrient molecule. 
The parameters were set as follows: $k=1000$, $m=10$, $\rho=0.075$, and $D=0.001$. 

{\bf Evolution}:
In each generation, a certain subset of cells with higher growth rates is selected to produce offspring, resulting in a given mutational change in the network. 
First, $n$ parent cells were generated, whose catalytic-reaction networks are randomly generated by means of connection rate $\rho$.
From each of the $n$ parent cells, $L$ mutant cells were generated by random replacement of $m \rho k^2$ reaction paths, where $\rho k^2$ is the total number of reactions, and $m$ denotes the mutation rate per reaction per generation.
To determine the growth rate of each cell, i.e., the inverse of the average time required for division, reaction dynamics were simulated for each of the $nL$ cells. Within the cell population, $n$ cells with faster growth rates were selected to be the parent cells of the next generation, from which $nL$ mutant cells were again generated in the same manner.
The parameters were set as follows: $n = 1000$, $L = 50$, and $m=1 \times 10^{-3}$. 
The simulation of evolutionary dynamics was performed under a constant (original) environmental condition ${\bf o}=\{s^o_1,\cdots,s^o_m \}$, where $s^o_1=s^o_2=\cdots =s^o_m=1/m$. 


\color{black}

\section{Global proportionality emerged through evolution}

Using the model described in the preceding section, we analyzed the response of the component concentrations to the environmental change from the original condition. Here, the environmental condition is given by the external concentration $s_1,\cdots,s_m$. We then changed the condition to $s^{(\varepsilon,{\bf E})}_j=(1-\varepsilon)s^o_j+\varepsilon s^{\bf E}_j$, where $\varepsilon$ is the intensity of the stressor and ${\bf E}=\{ s^{\bf E}_1,\cdots,s^{\bf E}_m \}$ denotes the vector of the new, stressed environment, in which the values of component $s^{\bf E}_1,\cdots,s^{\bf E}_m$ were determined randomly to satisfy $\sum_j s^{\bf E}_j =1$.
For each environment, we computed the reaction dynamics of the cell, and computed the concentration $x^{(\varepsilon,{\bf E} )}_j$ from the steady growth state to obtain the logarithmic change in the concentration $\delta X^{(\varepsilon,{\bf E} )}_j=\log (x^{(\varepsilon,{\bf E} )}_j/ x^{{\bf o}}_j)$; the change in growth rate $\mu$, designated as $\delta \mu^{(\varepsilon,{\bf E} )}$, was computed at the same time.

\color{black}
We next determined whether changes $\delta X^{(\varepsilon,{\bf E} )}_j$ satisfy the common proportionality trend across all components toward a variety of environmental changes, and tested whether the proportion coefficient is consistent with $\delta \mu^{(\varepsilon,{\bf E} )}$, relative to all the possible ranges in $0\leq \varepsilon \leq 1$. 
We examined the proportionality for the network both before and after evolution, i.e., the random network and the evolved network, under the given environmental condition.

\color{black}
First, we computed the response in expression to the same type of stress, i.e., the same vector ${\bf E}$ with different intensities $\varepsilon$. Fig. 2(a) shows the coefficients of correlation between the changes in component concentrations $\delta X^{(\varepsilon_1,{\bf E})}_j$ and $\delta X^{(\varepsilon_2,{\bf E})}_j$ caused by different magnitudes of environmental change ($\varepsilon_2 = \varepsilon_1 + \varepsilon$, at $\varepsilon >0$). This result was obtained using a variety of randomly chosen environmental vectors ${\bf E}$. As shown, for a small environmental change ($\varepsilon=0.02$), the correlation is sufficiently strong both for the random and evolved networks, whereas for a larger environmental change ($\varepsilon=0.08$), the correlation coefficients are significantly smaller for the random networks. Fig. 2(b) shows the relation between the ratio of the growth rate changes $\delta \mu^{(\varepsilon_1,{\bf E})}/\delta \mu^{(\varepsilon_2,{\bf E})}$ and the slope in $(\delta X^{(\varepsilon_1,{\bf E})}_j, \delta X^{(\varepsilon_2,{\bf E})}_j)$ obtained by fitting the concentration changes of all components. Fig. 2(c) shows the ratio of the slope to $\delta \mu^{(\varepsilon_1,{\bf E})}/\delta \mu^{(\varepsilon_2,{\bf E})}$ (which turns to unity when Eq. (1) is satisfied) as a function of the magnitude of environmental change $\varepsilon_1$. These results clearly showed that for a small environmental change, the relation in Eq. (1) is valid for both the random and evolved networks, while for the large environmental change, Eq. (1) holds only for the evolved networks.

\textcolor{black}{Next, we examined the correlation of concentration changes across different types of environmental stressors: the main topic of the present paper. Fig. 3(a) depicts examples of $(\delta X^{(\varepsilon,\bf{E_{\alpha}})}_j, \delta X^{(\varepsilon,\bf{E_{\beta}})}_j)$ obtained by three networks from different generations. For initial random networks, there is no correlation, whereas a modest correlation emerges in the 10th generation. Later, with evolution, common proportionality takes shape as seen for instance in the 150th generation in Fig. 3(a), where the proportionality reaches more than two digits. To demonstrate generality of the proportionality over a variety of environmental variations, we computed the coefficients of correlation between $\delta X^{(\varepsilon,\bf{E_{\alpha}})}_j$ and $\delta X^{(\varepsilon,\bf{E_{\beta}})}_j$ for a random choice of different vectors $\bf{E_{\alpha}}$ and $\bf{E_{\beta}}$. Fig. 3(b) shows the distributions of correlation coefficients obtained by the random network and evolved network (150th generation). Furthermore, Fig. 3(c) shows how they change as a function of generation: The correlation coefficients increase with generations, and those for the evolved network are much greater than coefficients in the initial random networks at generation$=0$ and are closer to unity. It is remarkable that global proportionality develops even for different environmental conditions that do not happen in the course of evolution.}

Next, the relation between $\delta \mu^{(\varepsilon_1,\bf{E_{\alpha}})}/\delta \mu^{(\varepsilon_2,\bf{E_{\beta}})}$ and the slope in $(\delta X^{(\varepsilon_1,\bf{E_{\alpha}})}_j, \delta X^{(\varepsilon_2,\bf{E_{\beta}})}_j)$ is presented in Fig. 3(d), while the ratio of the slope in $(\delta X^{(\varepsilon_1,\bf{E_{\alpha}})}_j, \delta X^{(\varepsilon_2,\bf{E_{\beta}})}_j)$ to $\delta \mu^{(\varepsilon_1,\bf{E_{\alpha}})}/\delta \mu^{(\varepsilon_2,\bf{E_{\beta}})}$ is shown in Fig. 3(e) as a function of $\varepsilon_2$. The agreement of the slope of $\delta X$ with the growth rate change given by Eq. (1) is much more remarkable for the evolved networks than for the random networks.

\textcolor{black}{
The global proportionality over all components across a variety of environmental conditions suggests that data $X^{(\varepsilon,{\bf E} )}_j$ across different environmental conditions are constrained to changes along a one-dimensional manifold after evolution (under a single environmental condition). To examine this possibility, we analyzed the changes in $X^{(\varepsilon,{\bf E} )}_j$ across a variety of environmental changes ${\bf E}$ and $\varepsilon$ by principal component analysis. Fig. 4 clearly reveals that in the evolved network, high-dimensional data from $X^{({\varepsilon,{\bf E}})}_j$ are located along a one-dimensional curve, while the principal-component plot is consistent with growth rate $\mu$ rather well (see Fig. 5). (Note that the contribution of the major-axis mode [i.e., the first principal component] reaches 74\% in the data.) In contrast, the data from the random network are scattered and no clear structure is visible.
}

After that, we examined the evolutionary course of the phenotype projected on the same principal-component space as that depicted in Fig. 6. As shown in Fig. 6(a), the points from $\{ X_j \}$ generated by random mutations to the reaction network are again located along the same one-dimensional curve. Furthermore, those obtained by environmental variation or via noise in the reaction dynamics also lie on this one-dimensional curve, as shown in Fig. 6(b). Hence, the phenotypic changes are highly restricted, both genetically and nongenetically, within an identical one-dimensional curve.

\color{black}
As shown in Figs. 4 and 6, the variation of concentration owing to perturbations is much larger along the first principal component than along other components. This finding suggests that relaxation is much slower in the direction of the first component than in other directions. To confirm that the first-principal-component mode has a predominantly slow time scale, we estimated the relaxation time of autocorrelation function for each principal component. To be specific, we computed the score of the $i$th principal component at the $j$th cell division $p^i_j$ during steady growth, and from the time series of component scores, we derived autocorrelation function $R^i(t)=\frac{1}{M} \sum_{j=1}^{M} p^i_j p^i_{j+t}$. Then, we estimated relaxation time $\tau_i$ by fitting the autocorrelation function to $R^i(t) = R^i_0 \exp (-t/\tau_i)$ with a constant, $R^i_0$. The time scales for each principal component are not substantially different for initial generations. As shown in Fig. 7, however, in the evolved network (i.e., generation=$120, 150$), the relaxation time of the first principal component is significantly larger than that of the others. This means that the fluctuation of cellular state under the influence of noise tends to be dominated by slow relaxation along the first component, whereas the fluctuations along orthogonal directions are relaxed much faster. 

\color{black}
In summary, we observed emergent global proportionality beyond trivial linearity in response to even tiny environmental changes.
After evolution, the linearity region expands to a level with an order-of-magnitude change in the growth rate. Notably, the proportionality over different components across different environmental conditions is enhanced through evolution. Changes in high-dimensional phenotype space across a variety of environmental conditions, genetic variations, and due to noise are nearly confined to a single one-dimensional manifold, i.e., the dimensionality in dynamics is strongly reduced from high-dimensional phenotypic space. This one-dimensional line given by the first principal-component axis highly correlates with the growth rate, and therefore the changes caused by environmental and evolutionary perturbations are constrained to changes along the same manifold.

\color{black}

\section{Dominance of the Major Axis Corresponding to the Growth Rate}

To formulate the observed global proportionality, we advanced the following hypothesis and discussed its consequences: 
Both environment- and evolution-induced changes in a high-dimensional phenotype space are constrained mainly to points near a one-dimensional major axis in principal component analysis \cite{footnote1}. Phenotypic dynamics are slower along this axis than across it. The axis forms a ridge in the fitness landscape. Evolution progresses along this axis, whereas the fitness function is steep across the axis. Indeed, dominance of the first principal mode in phenotypic change emerges after evolution simulation of the cell model (see Fig. 4).
Moreover, expression data from bacterial evolution studies support this hypothesis (see Fig. 6 in ref. \cite{CFKK-Interface}).

\color{black}
Although there is no mathematical derivation of this hypothesis, it may be plausible, considering the evolutionary robustness of phenotypes.
In phenotypic evolution, it is natural to assume that (1) the dynamics shaping phenotypes are high-dimensional and complex and are subject to stochastic perturbations, (2) dynamics (or genetic rules that govern them) shaping higher fitness states are rarer \cite{ESB}. Then, via evolution, dynamics leading to a higher fitness state will achieve robustness (resistance to perturbations), as discussed elsewhere \cite{Wagner,KK-PLoS}. Accordingly, in the state space, in order to shape phenotypes, there will be a flow from most directions to reach the selected phenotype, as shown schematically in Fig. 8. On the other hand, during the generations while the evolution is in progress, the states shift, thus yielding a further higher fitness state with mutational changes in the dynamics. Therefore, along the direction of such (mutational) perturbation of dynamics, phenotype states can be changed rather easily by perturbations. 
This scheme leads to the situation schematically shown in Fig. 8, i.e., only in the direction in which evolutionary change proceeds, is the relaxation slow, while for other directions orthogonal to it, change is much faster \cite{footnote2}. 

Now let us consider the relaxation dynamics to the attractor in a given generation. The magnitude of (negative) eigenvalues except for one (or a few) eigenvector(s) will be large, while that along the direction in which evolutionary change has occurred (and will proceed) will be much closer to zero, i.e., the relaxation in this direction is slower. Accordingly, the variance along the largest principal component will be dominant, as confirmed in Figs. 4 and 6 for numerical evolution of the model.

\color{black}

Now, by following the dynamical-system representation of phenotypic dynamics during steady growth \cite{Mu}, we will formulate the above hypothesis as follows. First, let us consider the high-dimensional phenotypic dynamics
\begin{equation}
dx_i/dt=f_i(\{x_j \})-\mu x_i
\end{equation}
where $x_i$ ($i=1,2,..,N$) is the concentration of components (e.g., proteins) in a cell, and $\mu$ is the growth rate of the cell. By means of $X_i=\log(x_i)$ and $x_iF_i=f_i$, the above equation is rewritten as

\begin{equation}
dX_i/dt=F_i(\{X_j \})-\mu.
\end{equation}
Thus, the stationary solution for a given environmental parameter vector ${\bf E}$ is expressed as
\begin{equation}
F_i(\{X_j^* \},{\bf E})=\mu ({\bf E }).
\end{equation}
By linearizing the equation around a given stationary solution using the Jacobi matrix ${\bf J}$ and susceptibility vector ${\bm \gamma}_i=\partial F_i/\partial {\bf E}$, we get 
\begin{equation}
\sum_j J_{ij}\delta X_j +{\bm \gamma}_i({\bf E}) \delta E=\delta\mu ({\bf E}).
\end{equation}

\color{black}
Thus, because of ${\bf L}={\bf J}^{-1}$,
\begin{equation}
{\bf \delta X} = {\bf L}(\delta\mu{\bf I}-{\bm \gamma} \delta E),
\end{equation} where ${\bf I}$ is a unit vector $(1,1,1,..1)^T$.

Let us denote eigenvalues of matrix ${\bf L}$ as $\lambda_k$, with the corresponding right (left) eigenvalues ${\bf w}_k$ (${\bf v}_k$), respectively. Then matrix $L$ is represented by $\sum_k \lambda_k {\bf w_k}{\bf v_k^T}$.

Hence, we get
\begin{equation}
{\bf \delta{X}} = \sum_k \lambda^k {\bf w_k} (\delta \mu ({\bf v_k^T\cdot}{\bf I})- ({\bf v_k^T\cdot}{\bm \gamma})\delta E ).
\end{equation}

The major-axis hypothesis postulates that the magnitude of the smallest eigenvalue of ${\bf J}$ (which is denoted as $k=0$)
is much smaller than that of the others (or the absolute eigenvalue of ${\bf L}={\bf J}^{-1}$ is much greater than that of the others), whose eigenmode dominates the long-term phenotypic dynamics, so that the change in ${\bf \delta X}$ is nearly confined to the major axis corresponding to its eigenvector (see the schematic in Fig. 8, and see Figs. 4--7 for possible numerical support).

Thus, to study the major response to environmental changes, only the largest eigenvalue of $L$ (i.e., the smallest eigenvalue of ${\bf J}$) with the corresponding eigenvector ${\bf w_0}$ is dominant.
The major change in the dynamics is projected onto this axis ${\bf w_0}$, as in the collective motion for a macroscopic variable in statistical physics \cite{Mori}.
In other words, the one-dimensional major axis along which ${\bf \delta X}$ are located, as uncovered in the preceding section (see Fig. 4) corresponds to ${\bf w_0}$. Using the reduction to this mode ${\bf w_0}$,  we get
\begin{equation}
{\bf \delta X} = \lambda^0 {\bf w_0}(\delta \mu ({\bf v_0 \cdot  I })-  ({\bf v_0 \cdot  \bm \gamma})\delta E). 
\end{equation}

Hence, across different environmental conditions, the following relation is obtained:
\begin{equation}
\frac{{\bf \delta X(E)}}{{\bf \delta X(E')}}= \frac{\delta\mu({\bf E})- {\bf v_0 \cdot}{\bm \gamma}({\bf E}) \delta E/({\bf v_0 \cdot}{\bf I})}{\delta\mu({\bf E'})- {\bf v_0 \cdot}{\bm \gamma} ({\bf E'})\delta E'/({\bf v_0 \cdot}{\bf I})}.
\end{equation}	

This relation results in proportionality of $\delta X$ across all components $j$. Reduction to the change along ${\bf w_0}$ explains the proportionality across all components, over different environmental conditions. Now, we will discuss the proportion coefficient. By setting \begin{math} \delta E=\delta \mu /\alpha ({\bf E})\end{math} assuming linearity, we get

\begin{equation}
\frac{{\bf \delta X(E)}}{{\bf \delta X(E')}}= \frac{\delta\mu({\bf E})}{\delta\mu({\bf E'})} 
\frac{(1- {\bf v_0 \cdot}{\bm \gamma}({\bf E})/(\alpha{\bf v_0 \cdot}{\bf I})}
{(1- {\bf v_0 \cdot}{\bm \gamma}({\bf E'})/(\alpha{\bf v_0 \cdot}{\bf I})} 
\end{equation}

Therefore, the proportion coefficient tightly correlates with $\delta \mu$, but there is a correction from $\delta \mu$ by the factor $(1- {\bf v_0 \cdot}{\bm \gamma}({\bf E})/(\alpha{\bf v_0 \cdot}{\bf I})) $. Nevertheless, this correction term would not be so large considering that the projection to the major axis is ${\bf w_0}$. Because the dynamics are robust from various directions except for major axis ${\bf w_0}$, perturbations to original dynamics except for this direction should decay rapidly. In other words, the change in the dynamics $d{\bf X}/dt={\bf F}({\bf X};{\bf E})$ caused by a perturbation may be large, so that the original state is regained (see Fig. 8). Then it is expected that ${\bm \gamma}$, sensitivity of ${\bf F}$ to an external change, takes a larger value for directions other than ${\bf w_0}$, i.e., ${\bm \gamma}$ vector consists mostly of components of ${\bf w_j}$ at $j\neq 0$. Given that ${\bf v_0\cdot w_j}=0$ for $j\neq 0$, this state of affairs suggests that $({\bf v_0 \cdot}{\bm \gamma})$ will be negligible. Hence, the correction term will be reduced to unity, independently of the direction of environmental perturbation. Thus, Eq. (1) follows: the proportion coefficient across different environmental conditions is approximately consistent with the ratio of the growth rates.

Here, we should note that under the assumption ${\bf v_0\cdot}{\bm \gamma} \sim 0$, we do not need the linearity between $\delta \mu$ and $\delta \epsilon$ to derive the linear relation between $\delta {\bf X}$ and $\delta \mu$. From Eq. (9), we directly obtain the relation
\begin{equation}
{\bf \delta X} = \lambda^0 \delta \mu {\bf w_0} ({\bf v_0\cdot I})
\end{equation}
without requiring the linear approximation $\delta \mu \propto \delta \varepsilon$.
In this scenario, even though $\delta \mu$ and ${\bf \delta X}$ show nonlinear dependence on $\delta \epsilon$, the agreement of the slope across the component changes $\delta X_j$ with the change in the growth rate $\delta \mu$ is derived. Indeed, the proportionality in the simulation results described in the last section holds up for the regime in which $\delta\mu$ is no longer proportional to stress intensity $\delta \epsilon$ (Fig. 9).

The above calculation based on linear algebra is also interpreted as follows (with possible nonlinear extension by means of the stable manifold).
Indeed, according to the numerical results in Figs. 4--7 and the schematic picture in Fig. 8, the change occurs along one-dimensional manifold ${\bf W}$, which corresponds to the first principal coordinate and is a nonlinear extension of ${\bf w_0}$, whereas the variation of the environmental change $\delta E$ is expressed in coordinate $W$ as $\delta W=W({\bf E}+\delta E)-W({\bf E})$. Then, after introduction of the projection from manifold ${\bf W}$ onto $X_j$ as $e^W_j$, the variation satisfies 
\begin{equation}
\delta X_j ({\bf E}) = e^W_j \delta W
\end{equation}
Because $e^W_j$ is determined only by the vector ${\bf W}$ and independent of environmental vectors ${\bf E}$, the global proportionality in $\delta X_j$ across components $j$ is obtained. Here, we should note that the change along ${\bf W}$ is parametrized by growth rate $\delta \mu$ because it has tight one-to-one correspondence with the principal coordinate. Then $\delta W$ is represented as a function of $f(\delta \mu)$. Given that $\delta W \propto \delta \mu ({\bf E})$ for small $\delta \mu$, linearization leads to Eq. (11).

\color{black}
To sum up, under the major-axis hypothesis, we could explain two basic features that have thus far been observed in experiments and simulations:

\noindent
 (1) {\bf Overall proportionality is observed in expression level changes across most components and across a variety of environmental conditions}. This is because high-dimensional changes are constrained to changes along the major axis, i.e., eigenvector ${\bf w_0}$.

\noindent
(2){\bf There is an extended region for global proportionality}. Because the range in the variation along  ${\bf w_0}$ is large, the change in phenotype is constrained to points near this eigenvector, so that the proportionality range of phenotypic change is extended via evolution. Furthermore, as long as the changes are nearly confined to the manifold along the major axis, global proportionality reaches the regime nonlinear to $\delta \epsilon$. 

\section{Consequences of the Theory for Phenotypic Evolution}

{\bf Congruence of phenotypic changes due to evolutionary and environmental changes}

 The phenotypic changes constrained to the small space around the major axis are shaped by evolution, which further constrains any subsequent evolutionary potential. 
Because both the environmentally and genetically induced changes of phenotype, $\delta {\bf X(E)}$ and $\delta {\bf X(G)}$, respectively,  are constrained to changes along the same axis, they inevitably correlate (recall Fig. 6).
Consequently, if we apply the argument for $\delta {\bf X(E)}$ to evolutionary (genetic) change $\delta {\bf X(G)}$, then the two changes
are given by 
\color{black}
\begin{equation}
{\bf J}{\bf \delta X}+{\bm \gamma}({\bf E}) {\delta E}+{\bm \gamma}({\bf G}) {\delta G}=\delta\mu ({\bf E}).
\end{equation}
Following the projection to the major axis again (and taking only the eigenmode), we get
\begin{equation}
\frac{{\bf \delta X}({\bf E})}{{\bf \delta X}({\bf G})}= \frac{\delta\mu ({\bf E})}{\delta\mu({\bf G})}\frac{(1- {\bf v_0\cdot}{\bm \gamma}({\bf E})/\alpha)}{ (1- {\bf v_0\cdot}{\bm \gamma}({\bf G})/\alpha)}.
\end{equation}
This reasoning shows proportionality between $\delta {\bf X}({\bf E})$ and $\delta {\bf X}({\bf G})$ across all components. As for the proportion coefficient, there may be a slight correction from $\delta \mu$ due to the factor $\frac{(1- {\bf v_0\cdot}{\bm \gamma}({\bf E})/\alpha)}{ (1- {\bf v_0\cdot}{\bm \gamma}({\bf G})/\alpha)}$. Nonetheless, this value is not expected to be so large as long as the projection of the ${\bm \gamma}$ vector to the major axis is not so large as already discussed in the preceding section (also recall Figs. 6 and 8, which show that the major phenotypic changes caused by environmental and genetic perturbations are located near the same manifold ${\bf W}$). Then, 
\color{black}
\begin{equation}
\frac{{\bf \delta X}({\bf E})}{{\bf \delta X}({\bf G})}= \frac{\delta\mu ({\bf E})}{\delta\mu({\bf G})} 
\end{equation} holds.
This proportionality between environmentally and evolutionarily induced changes, in the form of Eq. (15), is reported both in bacterial evolution experiments and numerical evolution of a cell model \cite{CFKK-Interface}; this pattern is well explained by the present theory. 

{\bf Fluctuation relation}

 Because changes due to noise and genetic variations progress along the common major axis, the phenotypic variances due to the former ($V_g$) and those due to the latter ($V_{ip}$) are proportional across all components. In fact, given that relaxation of phenotypic changes is much slower along major axis ${\bf W}$, the phenotypic fluctuations due to noise are nearly confined to this axis. Thus, the variance of each component $X_i$ is given by the means of the variance in variable $W$ of the principal mode (along the major axis) due to noise,
\color{black}
\begin{equation}
V_{ip}(i) =  (({\bf w_0})_i)^2 \langle \delta W^2 \rangle_\mathrm{noise}.
\end{equation}
Likewise, the variation due to the genetic change of each component is mostly constrained to points near the axis, so that
\begin{equation}
V_{g}(i) =  (({\bf w_0})_i)^2 \langle \delta W^2 \rangle_\mathrm{mutation},
\end{equation} where $({\bf w}_0)_i$ is the $i$th component of ${\bf w_0}$ (i.e., $\delta W ({\bf w_0})_i $ is in agreement with the projection of $\delta X_i$ onto the major axis ${\bf w^0}$).
Hence, \begin{equation}
V_{ip}(i) \propto V_{g}(i) 
\end{equation} holds across components.

Note that in the derivation of relation (18), the growth rate term is not involved. What we need is simply the dominance of the major-axis mode ${\bf w_0}$ in the expression changes. Then, the major variance in the high-dimensional expression change occurs along the single dimension ${\bf w_0}$, irrespectively of the source of perturbations, e.g., environmental change, noise, or genetic variation. Indeed, aside from the numerical evolution of the present cell model with the catalytic and reaction network \cite{CFKK-Interface}, this relation is also observed in the gene regulation network model where the growth rate dilution is not included \cite{ESB}, whereas some experimental data on {\sl Drosophila} \cite{Stearns} and yeast \cite{Landry, LehnerKK} support relation (18) at least partially.
The present theory accounts for the origin of the empirical relation as a direct consequence of the hypothesis about the dominance of major-axis mode, in which changes incurred both by (environmental) noise and (genetic) mutations are constrained to the small space around the axis. Thus, agreement with simulation results can provide further support for the hypothesis.

\color{black}

Here, recall that the evolutionary rate of a phenotype is proportional to its variance due to genetic change, $V_g$ \cite{Fisher,Hartl}. Hence, the $V_g-V_{ip}$ relation suggests that it is easier for a phenotype with larger $V_{ip}$ to evolve under selective pressure. Thus, the specific phenotype that is most likely to evolve is predetermined according to the variability due to noise prior to application of the mutation selection process. 

\section{Discussion}

In the present paper, we first confirmed the global proportionality across protein expression levels toward a variety of environmental conditions using the experimental data from a bacterial transcriptome and proteome. The rate of change in logarithmic concentrations is rather well consistent with that of the cellular growth rate. By carrying out an evolutionary simulation of a cell model with the catalytic reaction network, we showed that such proportionality is shaped through evolution, where expression changes in the high-dimensional state space of the phenotype are constrained to changes along a major axis that corresponds to the growth rate.

These results are explained by the assumption that most of the expression changes are constrained to points near the major axis ${\bf w_0}$.
This axis ${\bf w_0}$ is the eigenvector for the smallest eigenvalue of the Jacobi matrix in the expression reaction dynamics; accordingly, the variation in phenotype ${\bf \delta X}$ is larger along this axis. Biologically speaking, this ${\bf w_0}$ might be represented by a concentration change of a specific component, but more naturally, this axis could be represented by a collective variable generated from the concentrations (expression levels) of many components. In other words, changes of the cellular state that progresses in a high-dimensional space are nearly confined to changes along this major axis ${\bf w_0}$. 

\textcolor{black}{Note that once the dominance in major axis ${\bf w_0}$ is shaped by evolution, linearization against environmental condition $\delta {\bf E}$ is no longer necessary. Furthermore, as long as the dominance is shaped through evolution under given fitness \cite{ESB}, the global proportionality holds even if fitness is not determined by the growth rate. In fact, using the cell model that we described here but selecting higher concentration of a given target chemical as the fitness for selection (instead of the growth rate), we again observed global proportionality over all components after evolution. In fact, as shown in Fig. S2(a), the correlation of concentration changes across different types of environmental changes increases with generations, indicating that global
proportionality for different environmental conditions is achieved. 
After the evolution, the slope in $(\delta X^{(\varepsilon_1,\bf{E_{\alpha}})}_j, \delta X^{(\varepsilon_2,\bf{E_{\beta}})}_j)$ (with random selection of different vectors $\bf{E_{\alpha}}$ and $\bf{E_{\beta}}$ and magnitudes $\varepsilon_1$ and $\varepsilon_2$) is in good agreement with the ratio of fitness change $\delta f^{(\varepsilon_1,\bf{E_{\alpha}})}/\delta f^{(\varepsilon_2,\bf{E_{\beta}})}$, where $\delta f^{(\varepsilon,\bf{E})}$ indicates the change of the fitness (i.e., the concentration of target chemical) observed in the corresponding environment (Fig. S2(b)). The principal component analysis also supported the notion that concentration changes caused by random environmental perturbations are confined to a low-dimensional curve (Fig. S2(c)), where the first principal component corresponds to fitness (Fig. S2(d)). These results in Fig. S2 strongly indicate that the arguments in the present paper are valid during evolution of given fitness.}

Because the phenotypic change due to noise or environmental or evolutionary change is mainly constrained and larger along this major axis ${\bf w_0}$, the phenotypic properties along this axis  can be considered plastic (see Figs. 4 and 6), while the robustness of fitness (resistance to mutation and noise) develops \cite{Wagner,KK-PLoS,ESB} in the course of evolution under a fixed condition, as discussed in the nearly-neutral theory by Ohta \cite{Ohta}. 
The environmental or genetic perturbation is buffered to the change along the major axis, ${\bf \delta X \cdot w_0}$, while the fitness (growth rate) change achieves robustness after evolution. Indeed, the compatibility of plasticity with robustness remains one of the major issues in biology \cite{Waddington,deVisser,Callahan,Pigliucci,Kirschner,Hatakeyama}. 

The question remains as to why such an axis, along which the change is much larger than that occurring in other directions, is shaped by evolution. 
As already discussed, robustness is shaped by evolution, so that stronger attraction is achieved in the state space for most directions. 
By contrast, along the axis in which phenotypes change via evolution, plasticity, i.e., changeability, remains. 
Consequently, we argued that phenotype dynamics in various directions except for those relevant to the evolutionary changes 
are expected to undergo strong contraction as evolution progresses. This result leads to the dominance of change along the major axis, as discussed here.

Note that the dominance of changes along the major axis, once it is shaped by evolution, can facilitate further evolution. If there were no such directions, and instead random mutations in many directions were the main source of phenotypic change, the changes would mostly cancel each other out, so that the phenotypic changes corresponding to specific fitness growth would be suppressed, resulting in the reduction of evolvability. Instead, if all the changes are confined to the mode along the major axis ${\bf w_0}$, then a certain degree of phenotypic change would be assured to foster subsequent evolution. 

In fact, the variation in $X$ due to environmental or evolutionary change remains rather large.	Both the principal component analyses of bacterial evolution \cite{CFKK-Interface} and the results of the present study suggest that the evolutionary course in gene expression patterns can be described by one (or a few) major degrees of freedom, and 
this ``major axis'' extracted by the principal component analysis highly correlates with the growth rate of the cell. 

\textcolor{black}{
In the present paper, we discuss the emergence of a dominant mode of evolution in a fixed environment. One might wonder whether such single mode forms if the evolution progresses during fluctuation across several environmental or fitness conditions. In this case, one would expect, that not only a single mode but multiple principal components would make a larger contribution through evolution. For example, when we consider evolution under the influence of a variety of antibiotics, reduction to a single dominant mode may be too simple. Still, we naturally expect the reduction to a few modes, in accordance with the notion of robustness shaped by evolution. Indeed, one study on antibiotic-resistant {\sl E. coli} strains revealed that the resistance to 26 antibiotics via various mechanisms of action can be quantitatively represented by 7 or 8 modes represented by expression levels of a few thousand genes \cite{Suzuki}. This result suggests that there is a small number of major modes that govern high-dimensional expression dynamics leading to antibiotic resistance.}

\textcolor{black}{
Formation of such low-dimensional structure from a complex high-dimensional state space was recently discussed in terms of a variety of biological problems, such as neural-network dynamics \cite{Brenner}, protein dynamics \cite{Tlusty}, and laboratory ecological evolution \cite{Leibler}, whereas separation of a slow eigenmode has also been discussed for biological data \cite{sloppy} and simulations \cite{Kohso}. Formation of a dominant mode through evolutionary robustness will be important for discussing universality of such a dimension reduction in biological systems.}

To sum up, the present study opens up the possibility to describe adaptation and evolution in response to environmental changes by only a few collective variables. By means of this representation, evolvability, i.e., the changeability of a phenotype via evolution, can be quantitatively characterized. This changeability is also represented by phenotypic fluctuations due to noise or environmental variation, as in the evolutionary fluctuation--response relation \cite{Sato,ESB}. In other words, the phenotype that is more evolvable is predetermined before a genetic change occurs; this mechanism is a manifestation of the genetic assimilation concept proposed by Waddington \cite{Waddington}.

Of course, further studies are needed to confirm the generality of the evolution of dominance of major-axis mode, both in experiments and numerical simulations.
\textcolor{black}{As discussed above, during evolution in terms of given fitness, which is not necessarily the growth rate, dominance of the major-axis mode is shaped, thereby leading to the global proportionality of concentration changes (see Fig. S2).
 Moreover, we previously studied gene regulation networks in which the expression of many genes mutually activates or suppresses the expression of other genes \cite{KK-PLoS,ESB}.
 Preliminary results suggest that global proportionality is a general property of evolved networks; that is, all expression levels change in proportion up to a large change in fitness, and the expression changes are located along a one-dimensional manifold, as represented by the principal component that correlates with fitness.
In this case again, the dominance in the major-axis (or slow-manifold) mode itself seems to be a general consequence of evolution toward robustness \cite{KK-PLoS}, even without the use of growth rate $\mu$ as fitness. 
When global dilution by the growth rate is imposed and fitness correlates with the growth rate, proportionality of the component with $\delta \mu$ is achieved.}
 
Another important future extension of our theory will be formulation 
not associated with the steady growth state. In cells, there are states in which growth is suppressed, such as the stationary state \cite{Monod,stationary2,Balaban}, and some theoretical approaches have been proposed to describe these nongrowth states \cite{Dill}.
It will be worthwhile to determine whether (or how) the incorporation of more degrees of freedom in addition to the major axis can capture the state with $\mu \sim 0$. 

~\\
\noindent
{\bf Acknowledgment}

The authors would like to thank Matthias Heinemann for useful discussions of his experiment.
The authors are grateful to Tetsuhiro Hatakeyama, Pablo Sartori, and Bingkan Xue for stimulating discussions.
This research was partially supported by the Platform for Dynamic Approaches to Living Systems from the Japan Agency for Medical Research and Development (AMED) [to K.K], Grant-in-Aid for Scientific Research (S) (15H05746 [to K.K. and C.F]), and by Grant-in-Aid for Scientific Research (B) (15H04733, 15KT0085 [to CF]) from the Japan Society for the Promotion of Science (JSPS). KK would also like to acknowledge Charles L. Brown Membership at the Institute for Advanced Study at Princeton.

~\\
~\\
~\\
\noindent
{\Large {\bf Figure captions}}\\
~\\
{\bf Figure 1.} Common proportionality in {\it E. coli} protein expression changes. 
Panels (a) and (b) show examples of the relation between protein expression changes $\delta X_j({\bf E_{\alpha}})$ and $\delta X_j({\bf E_{\beta}})$. $\delta X_j({\bf E})$ represents the log-transformed expression changes of the $j$th protein between environment ${\bf E}$ and the standard condition. 
\textcolor{black}{Here, we chose the M9 minimal medium with glucose (as the carbon source) as the standard condition. (a) The relation between the glycerol carbon source condition and chemostat culture with $\mu=0.12 \mathrm{h}^{-1}$, and panel (b) shows the relation between the mannose carbon source and acetate carbon source conditions. The blue line represents the slope calculated by the ratio of observed growth rate changes. In this analysis, we used the protein expression data of genes with expression levels in all environments that exceed the threshold ($x_j({\bf E})>50$), to exclude any inaccurate data ($\sim$40\% of the expression data were discarded). (c) Relation between the slope of the change in protein expression and the change in the growth rate. The abscissa represents $\delta\mu({\bf E_1}) / \delta\mu({\bf E_2})$, while the ordinate is the slope in $\delta X_j({\bf E_1}) / \delta X_j({\bf E_2})$. The slope was obtained by fitting the protein expression data by the major-axis method, which is a least-squares fit method that treats the horizontal and vertical axes equally and is commonly used to fit bivariate scatter data \cite{Fitting_Method}. The pairs ${\bf E_1}$ and ${\bf E_2}$ satisfying $\mu({\bf E_2})<\mu({\bf E_1})<\mu({\bf S})$ were plotted, where $\mu({\bf S})$ represents the growth rate under the standard condition. The relation obtained by choosing the standard condition from all possible data is presented in Supplemental Fig. S1.\\}
~\\
{\bf Figure 2.} Common proportionality in concentration changes in response to the same type of stress in the catalytic-reaction network model. 
(a) The coefficient of correlation between the changes in component concentrations $\delta X^{(\varepsilon_1,{\bf E})}_j$ and $\delta X^{(\varepsilon_2,{\bf E})}_j$ with $\varepsilon_2 = \varepsilon_1 + \varepsilon$. For the random and evolved networks, the correlation coefficients with a small ($\varepsilon=0.02$) and large ($\varepsilon=0.08$) environmental change are plotted, which were obtained using 100 randomly chosen environmental vectors ${\bf E}$. Throughout the study, each component concentration $\delta X_j$ was calculated by averaging 100 cell division events after the cell reaches a steady-growth state, except for the results shown in Figs. 6(b) and 7. 
(b) Relation between the ratio of growth rate change $\delta \mu^{(\varepsilon_1,{\bf E})}/\delta \mu^{(\varepsilon_2,{\bf E})}$ and the slope in $(\delta X^{(\varepsilon_1,{\bf E})}_j, \delta X^{(\varepsilon_2,{\bf E})}_j)$ obtained by fitting the concentration changes of all components. 
(c) The ratio of the slope in the concentration changes to the growth rate change as a function of the intensity of stress $\varepsilon$. The ratio in the ordinate becomes unity when Eq. (1) is satisfied.\\
~\\
{\bf Figure 3.} Common proportionality in concentration changes in response to different types of stress, in the catalytic-reaction network model. 
\textcolor{black}{(a) Concentration changes across different types of environmental stressors. 
For three different networks from different generations, $(\delta X_j(\bf{E_{\alpha}}), \delta X_j(\bf{E_{\beta}}))$ are plotted by means of different randomly chosen vectors $\bf{E_{\alpha}}$ and $\bf{E_{\beta}}$.
(b) Distributions of coefficients of correlation between the changes in component concentrations $\delta X^{(\varepsilon,{\bf E_{\alpha}})}_j$ and $\delta X^{(\varepsilon,{\bf E_{\beta}})}_j$. Red and green curves represent the distributions of a random network and evolved network (150th generation) obtained from 1000 pairs of ${\bf E_{\alpha}}$ and ${\bf E_{\beta}}$, respectively. The magnitude of environmental change $\varepsilon$ is fixed at $0.8$.
(c) The median of correlation coefficients as a function of generation. For each generation, we randomly selected a network from the population and the correlation coefficients were obtained using 100 randomly chosen pairs of two environmental vectors ${\bf E_{\alpha}}$ and ${\bf E_{\beta}}$. The magnitude of environmental change $\varepsilon$ is fixed at $0.8$. The bars represent the interquartile range of correlation coefficients. 
(d) Relation between the ratio of growth rate change $\delta \mu^{(\varepsilon_1,{\bf E_{\alpha}})}/\delta \mu^{(\varepsilon_2,{\bf E_{\beta}})}$ and the slope in $(\delta X^{(\varepsilon_1,{\bf E_{\alpha}})}_j, \delta X^{(\varepsilon_2,{\bf E_{\beta}})}_j)$. 
(e) The ratio of the slope in the relation between concentration changes and growth rate change as a function of the intensity of stress $\varepsilon$ in the case of different types of stress. }\\
~\\
{\bf Figure 4.} The change of $X^{(\varepsilon,{\bf E})}_j$ with environmental changes in principal-component space. 
Component concentrations $X^{(\varepsilon,{\bf E})}_j$ at randomly chosen various ${\bf E}$ and $\varepsilon$ values are presented for (a) evolved and (b) random networks. Here, the contribution of the first, second, and third components is 74\%, 8\%, and 5\%, respectively. \\
~\\
{\bf Figure 5.} The growth rate plotted against the first principal components by means of the computation and data from Fig. 4(a).\\
~\\
{\bf Figure 6.} A change in component concentration $X_j$ because of (a) mutations and (b) noise in reaction dynamics. In (a), mutations were added to the evolved reaction network by randomly replacing 0.5\% of the reaction paths. The red dots show the concentrations of components after mutations, which are projected onto the same principal-component space as that depicted in Fig. 4(a). The gray dots represent the concentration changes caused by environmental changes for the reference, which are identical to those shown in Fig. 4(a). The red dots in (b) represent the concentration changes observed at each cell division, which are caused by the stochastic nature of reaction dynamics. These data (as well as those in Fig. 4) were obtained under the original environmental condition employed for the evolutionary simulation. \\
~\\
\textcolor{black}{
{\bf Figure 7.} Relaxation time along with a principal component. A network was randomly selected from populations of generations $0, 30, 60, 90, 120$, and $150$. Next, the autocorrelation function of the first 5 principal-component scores $p^i_j$ for component $i$ at cell division $j$ was calculated from the concentration changes over 500 sequential cell divisions in the environment that was chosen for the evolutionary simulation. Autocorrelation function $R^i(t)=\frac{1}{M} \sum_{j=1}^{M} p^i_j p^i_{j+t}$ was computed from the data, and relaxation time $\tau_i$ was estimated by fitting the autocorrelation function to $R^i(t) = R_0 \exp (-t/\tau_i)$. Relaxation time $\tau_i$ for component $i=1,2,..,5$ is plotted for the 6 generations shown. \\}
~\\
{\bf Figure 8.} Schematic representation of the major-axis hypothesis. In the state space of ${\bf X}$, dominant changes are constrained to points near the ${\bf w^0}$ axis (and its connected manifold), while the attraction to this manifold is faster.\\
~\\
{\bf Figure 9.} Change in $X_i$ plotted against the change in $\mu$ (a) and against $\varepsilon$ (b) in the cell model. Only a few components $i$ among 1000 are plotted for the evolved network. The linear change is more extended in the plot against $\mu$ than in that against $\varepsilon$.\\
~\\
{\bf Supplemental Figure S1.}
\textcolor{black}{The relation between the slope of the change in protein expression and the change in growth rate obtained by choosing all possible environmental conditions as standard condition $S$. As in Fig. 1(c), The abscissa represents $\delta\mu({\bf E_1}) / \delta\mu({\bf E_2})$, while the ordinate is the slope in $\delta X_j({\bf E_1}) / \delta X_j({\bf E_2})$. The pairs ${\bf E_1}$ and ${\bf E_2}$ satisfying $\mu({\bf E_2})<\mu({\bf E_1})<\mu({\bf S})$ were plotted, where $\mu({\bf S})$ represents the growth rate under the standard condition.}\\
~\\
{\bf Supplemental Figure S2.}
\textcolor{black}{Results of evolution via selection of cells with higher target concentrations. In this evolutionary simulation, cells with a higher concentration of a target chemical are selected to be the parent cells of the next generation. Other parameters are identical to those in Figs. 2--7.
(a) The average of coefficients of correlation between the changes in component concentrations $\delta X^{(\varepsilon,{\bf E_{\alpha}})}_j$ and $\delta X^{(\varepsilon,{\bf E_{\beta}})}_j$, as a function of generation. For each generation, we randomly selected a network from the population and the correlation coefficients were obtained using 100 randomly chosen pairs of two environmental vectors ${\bf E_{\alpha}}$ and ${\bf E_{\beta}}$. The magnitude of environmental change $\varepsilon$ is fixed at $0.8$. The error bars represent the standard deviation of correlation coefficients. 
(b) The relation between the ratio of fitness change $\delta f^{(\varepsilon_1,{\bf E_{\alpha}})}/\delta f^{(\varepsilon_2,{\bf E_{\beta}})}$ and the slope in $(\delta X^{(\varepsilon_1,{\bf E_{\alpha}})}_j, \delta X^{(\varepsilon_2,{\bf E_{\beta}})}_j)$. Here, the fitness $f^{(\varepsilon,{\bf E})}$ is the concentration of the target chemical in the given environment. 
(c) The change of $X^{(\varepsilon,{\bf E})}_j$ with environmental changes in the principal-component space. The component concentrations $X^{(\varepsilon,{\bf E})}_j$ at a variety of randomly chosen ${\bf E}$ and $\varepsilon$ values are presented for an evolved network (in the 120th generation), where the contribution of the first, second, and third components is 40\%, 16\%, and 8\%, respectively. 
(d) The growth rate plotted against the first principal components on the basis of the computation and data from Fig. S2(c).}

\newpage

\begin{figure}[h]
\begin{center}
\includegraphics[width=12cm]{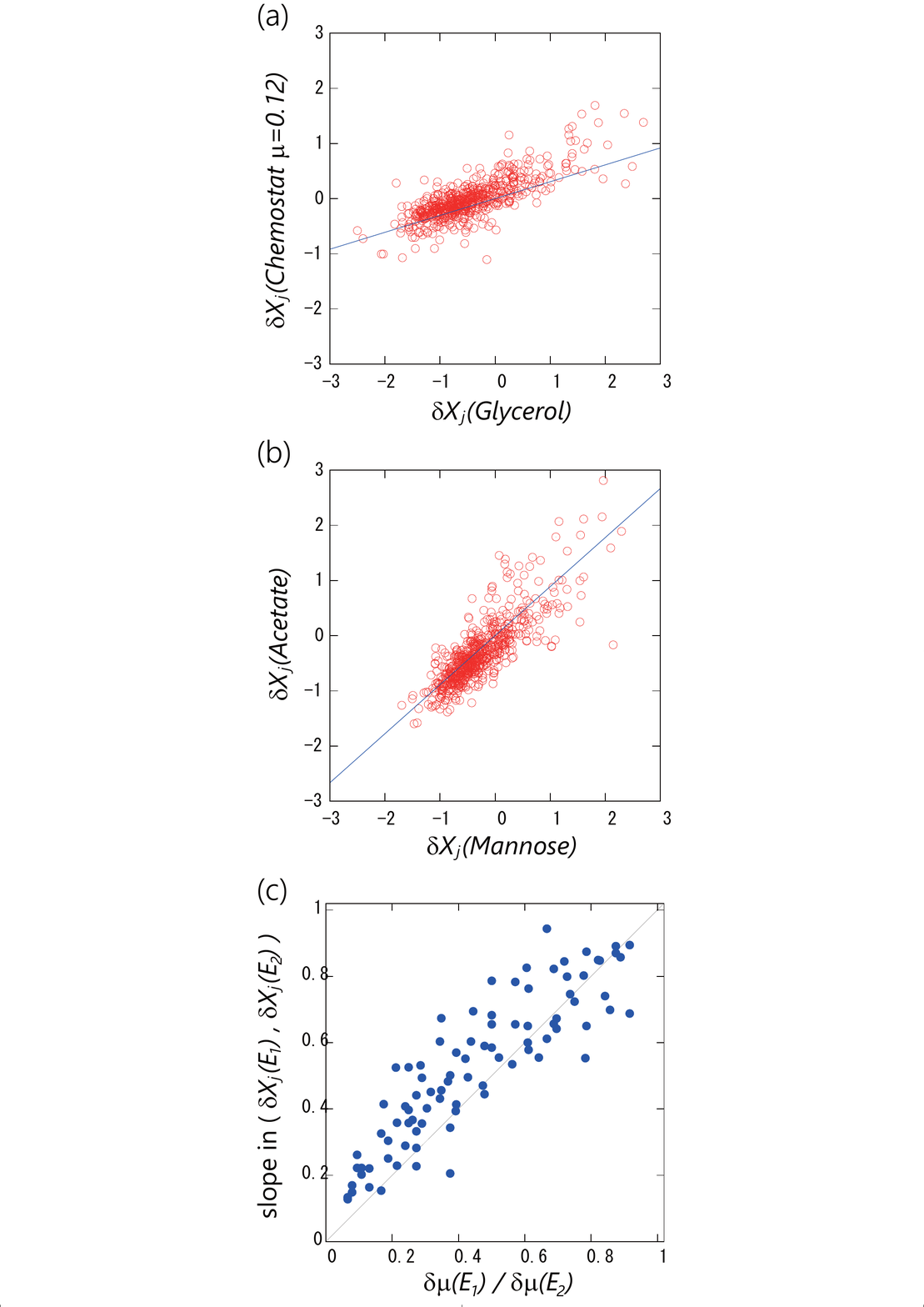}
\end{center}
\caption{
}
\end{figure}

\newpage
\begin{figure}[h]
\begin{center}
\includegraphics[width=12cm]{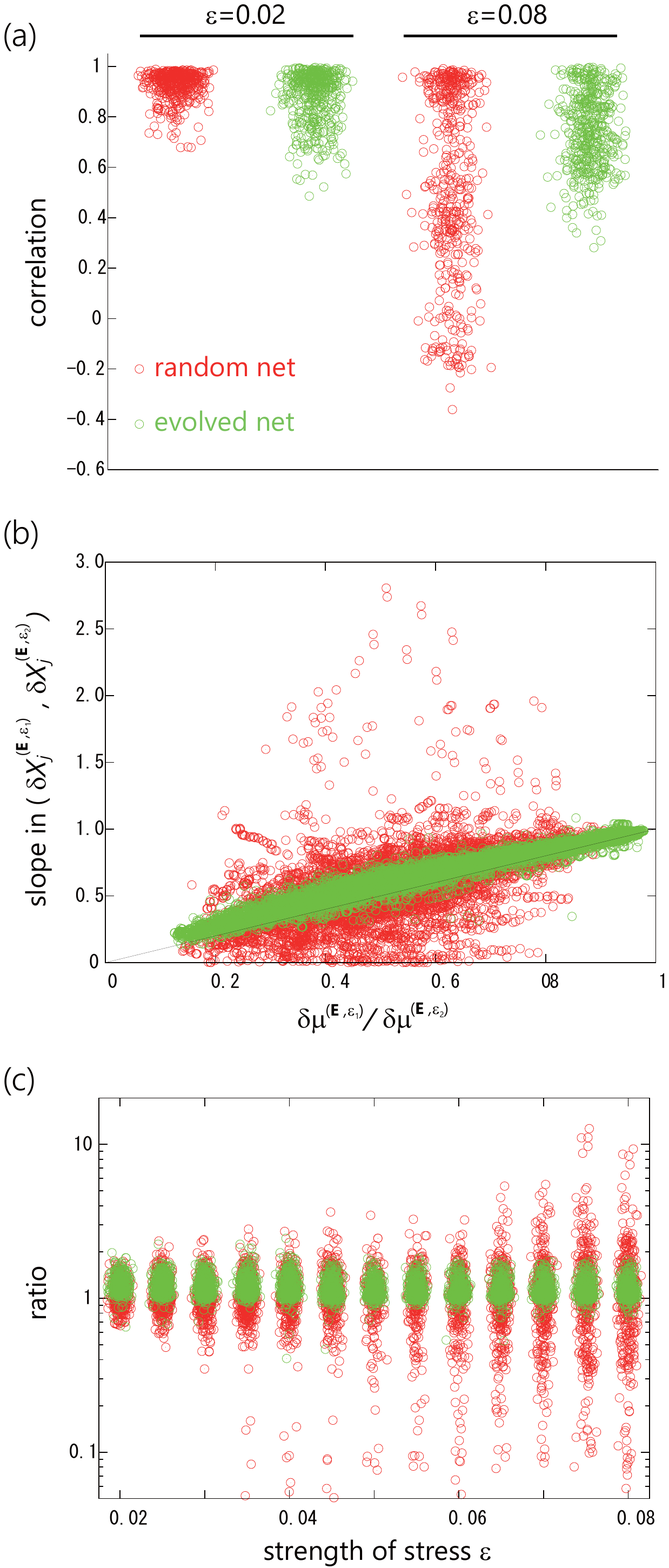}
\end{center}
\caption{
}
\end{figure}

\newpage
\begin{figure}[h]
\begin{center}
\includegraphics[width=16cm]{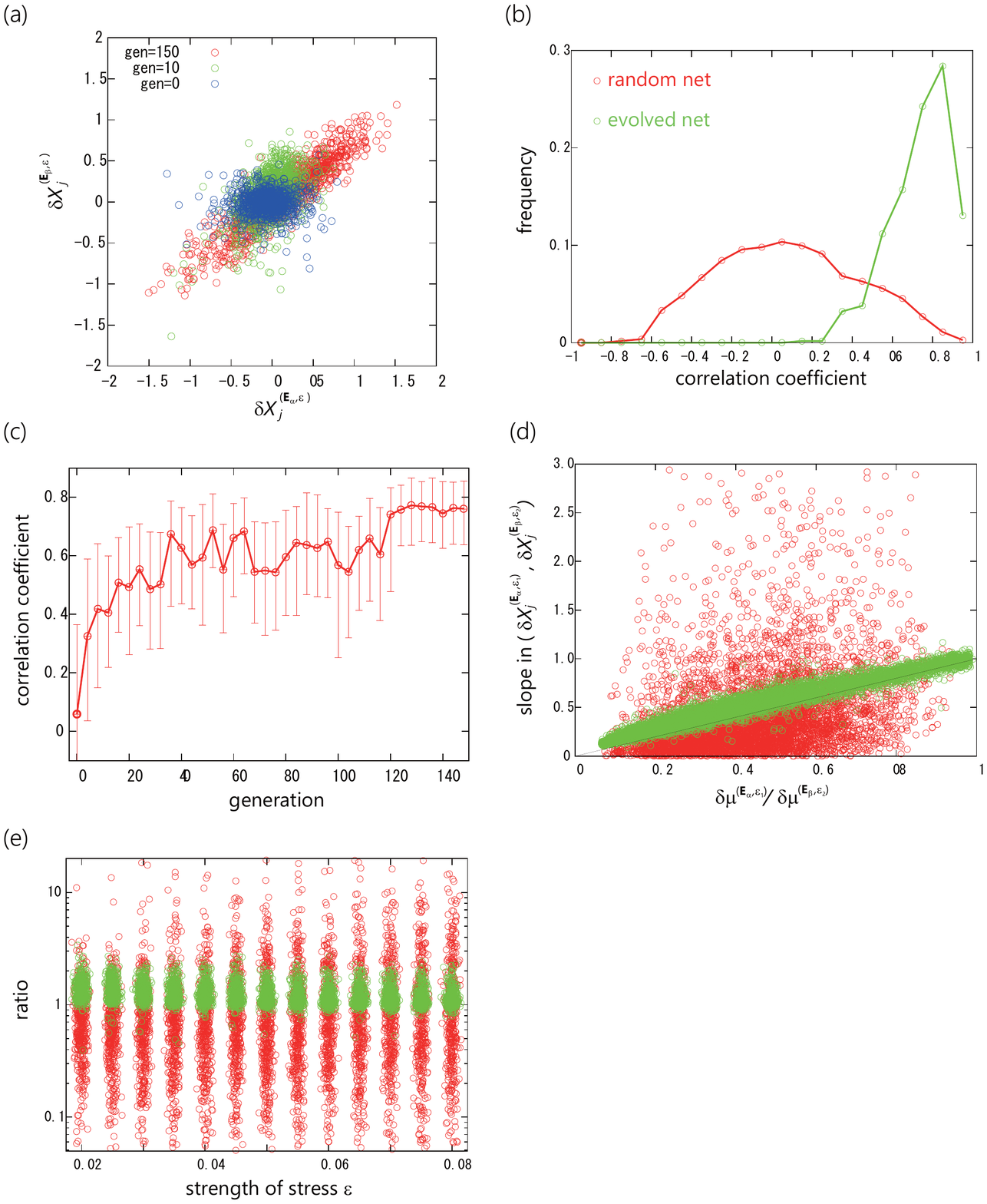}
\end{center}
\caption{
}
\end{figure}

\newpage
\begin{figure}[h]
\begin{center}
\includegraphics[width=12cm]{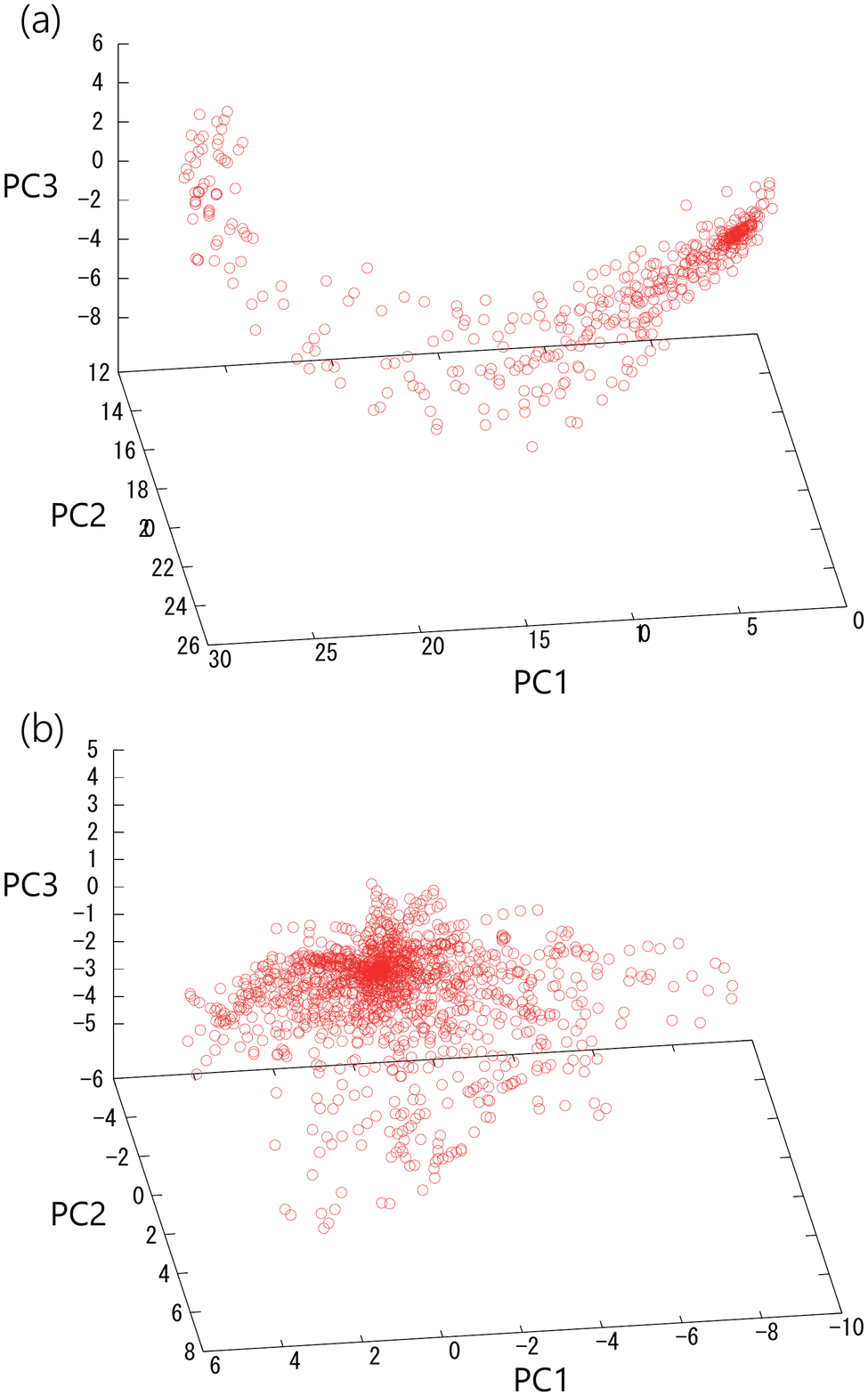}
\end{center}
\caption{
}
\end{figure}

\newpage
\begin{figure}[h]
\begin{center}
\includegraphics[width=12cm]{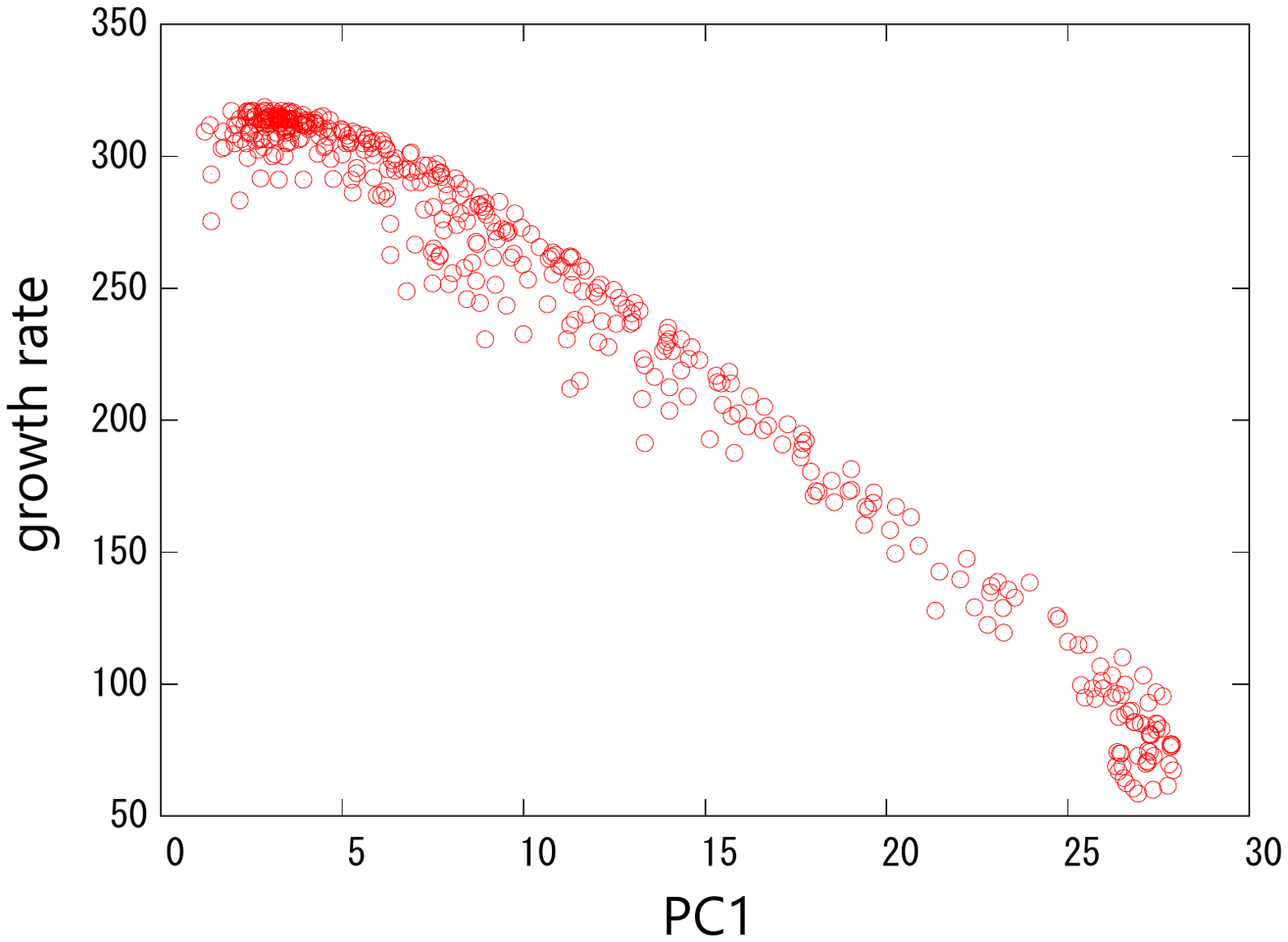}
\end{center}
\caption{
}
\end{figure}

\newpage
\begin{figure}[h]
\begin{center}
\includegraphics[width=12cm]{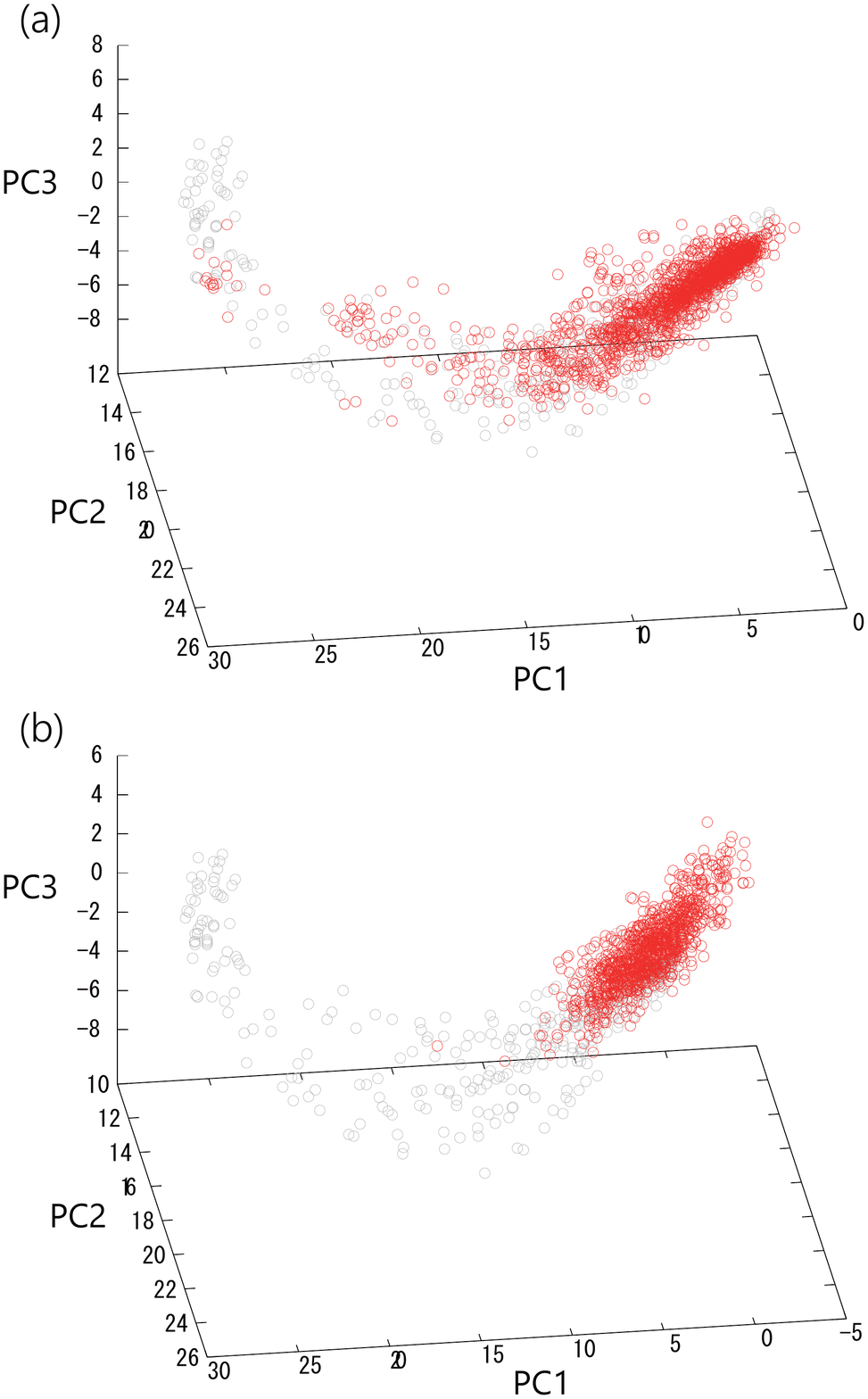}
\end{center}
\caption{
}
\end{figure}

\vspace{10cm}
\begin{figure}[h]
\begin{center}
\includegraphics[width=12cm]{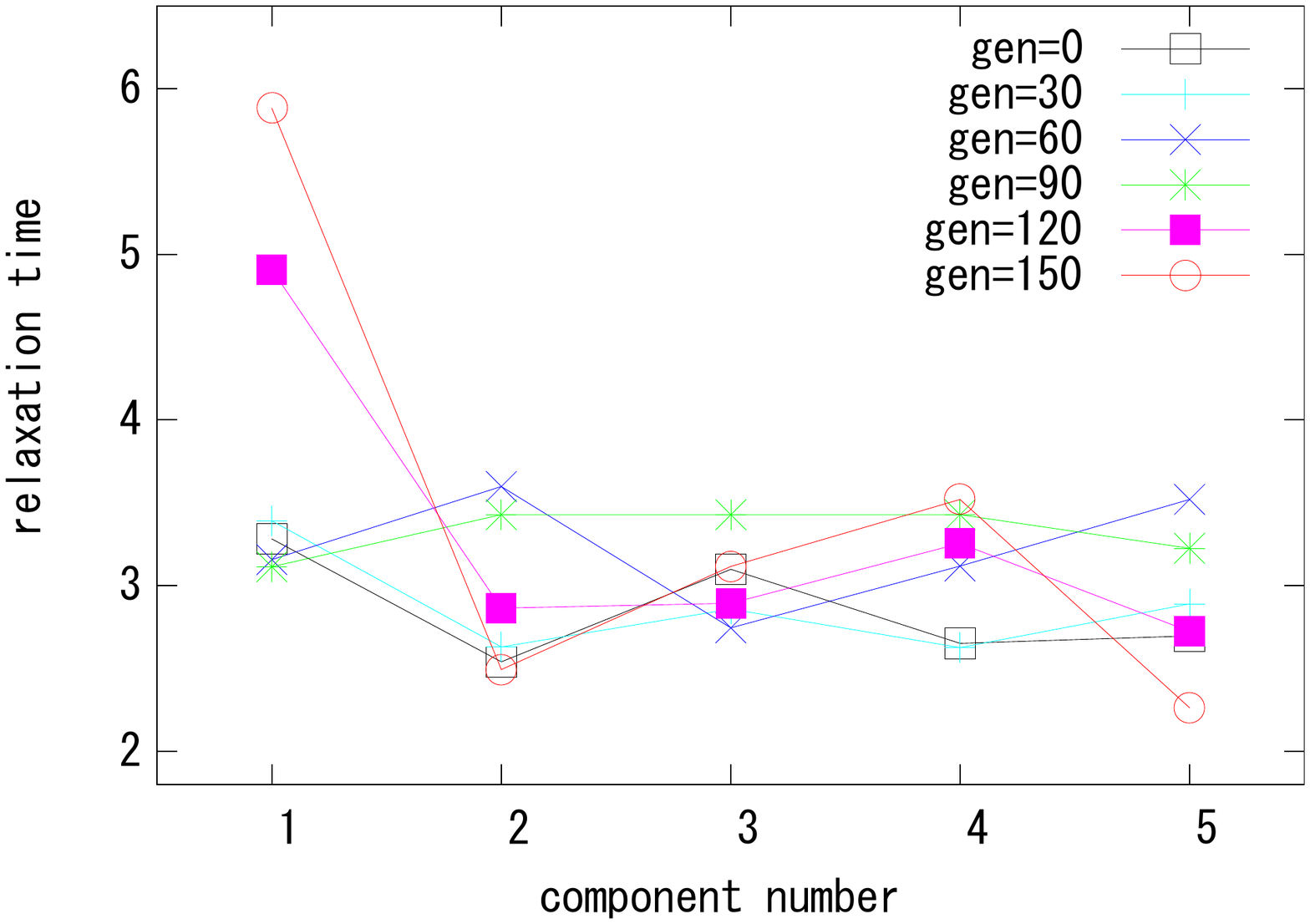}
\end{center}
\caption{
}
\end{figure}

\begin{figure}[h]
\begin{center}
\includegraphics[width=12cm]{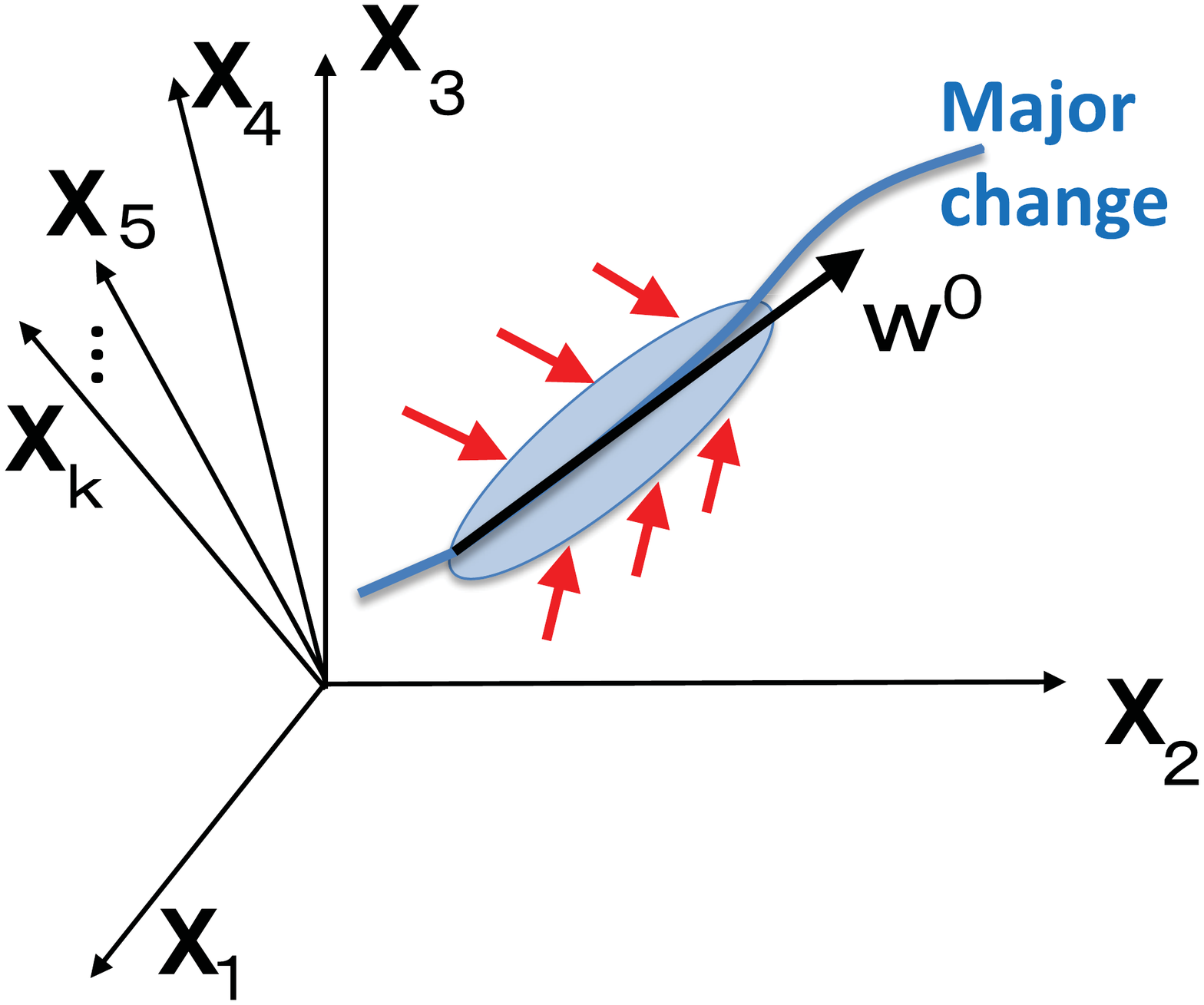}
\end{center}
\caption{
}
\end{figure}

\begin{figure}[h]
\begin{center}
\includegraphics[width=12cm]{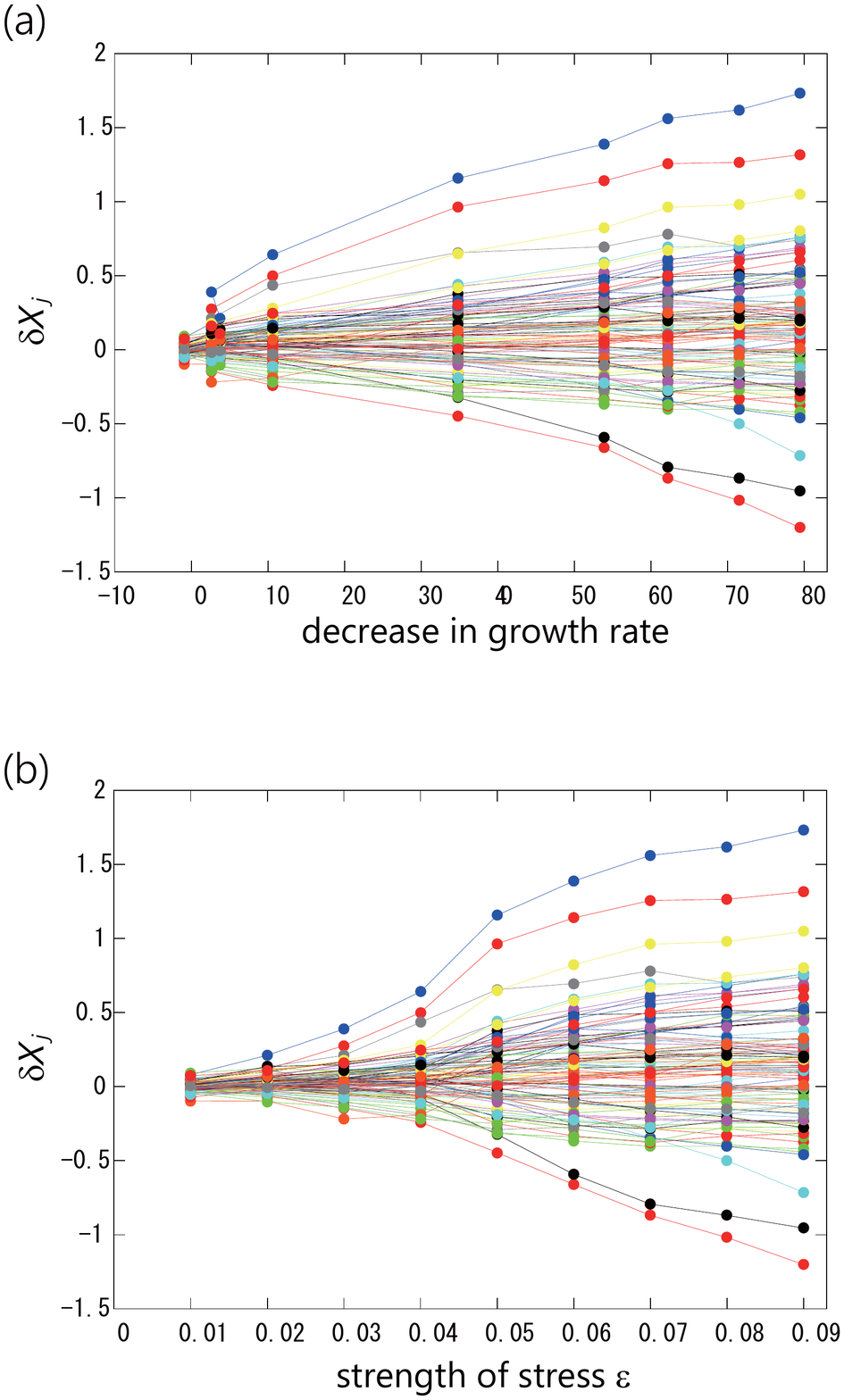}
\end{center}
\caption{
}
\end{figure}

\renewcommand{\figurename}{\bf Supplemental Figure S}

\begin{figure}[h]
\begin{center}
\includegraphics[width=12cm]{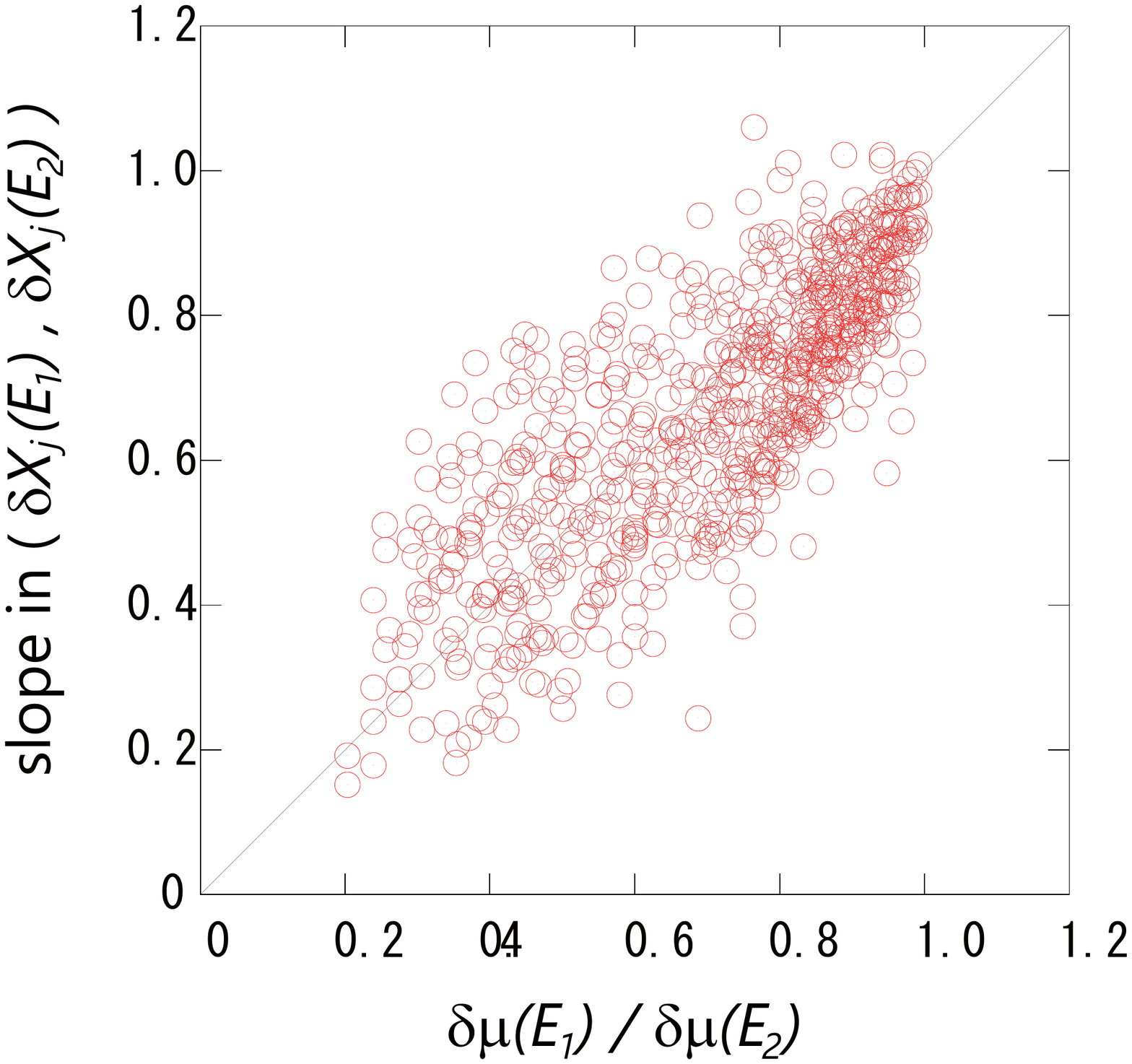}
\end{center}
\captionsetup{labelformat=empty,labelsep=none}
\caption{
{\bf Supplemental Figure S1}
}
\end{figure}

\begin{figure}[h]
\begin{center}
\includegraphics[width=16cm]{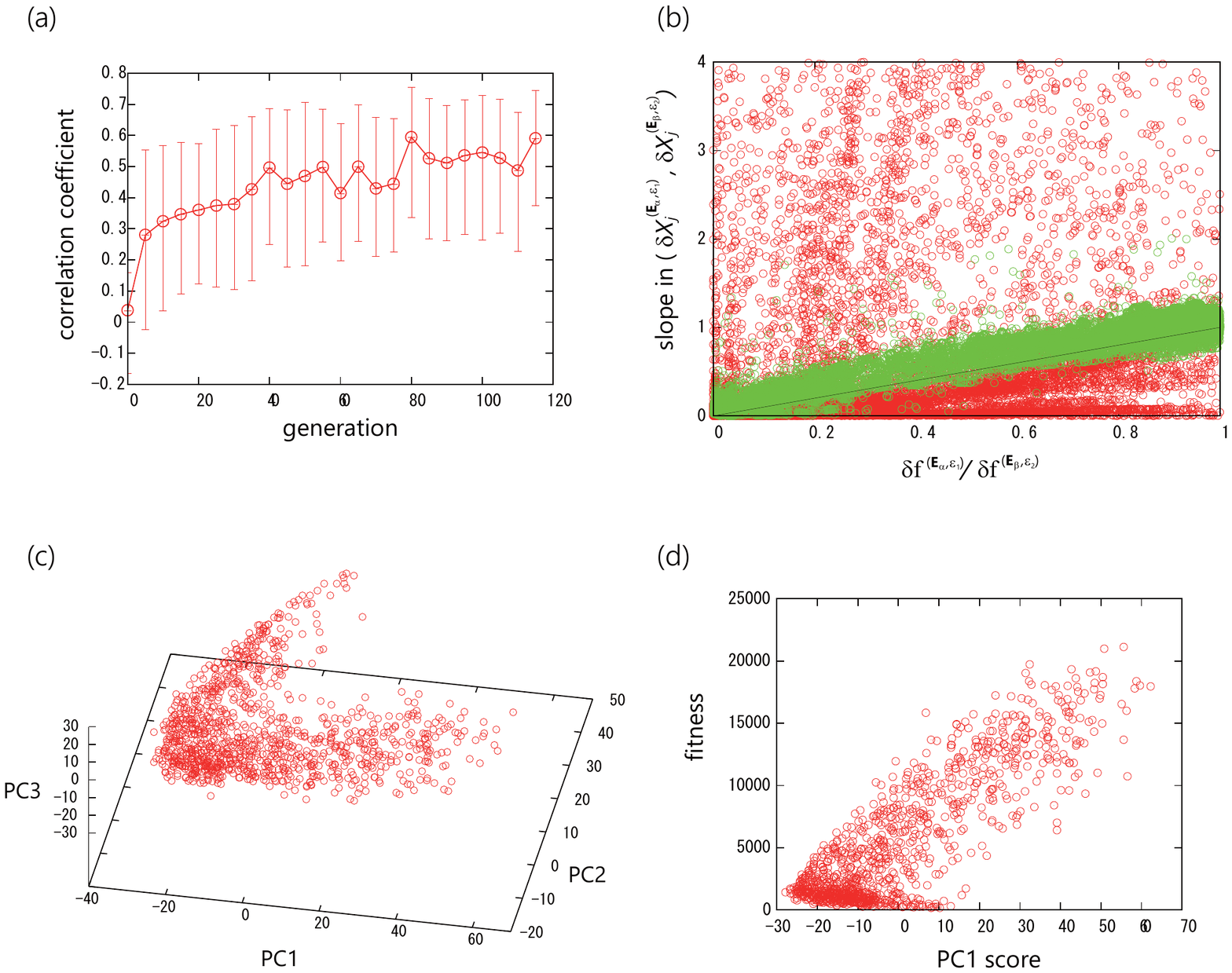}
\end{center}
\captionsetup{labelformat=empty,labelsep=none}
\caption{
{\bf Supplemental Figure S2}
}
\end{figure}


\begin{thebibliography}{999}
\bibitem{footnote1}
Mathematically, it would be represented as a slow manifold.
\bibitem{footnote2}
There may be the case in which there are more than one direction close to marginal stability. Especially, this may be the case when evolution reaches the fittest state and neutral space is shaped \cite{Wagner}. Even in this case, reduction to a few degrees of freedom is expected. Here, we are mainly concerned with the case when evolution is ongoing, and only one dimension may remain.
\end{thebibliography}
\end{document}